\documentclass[twocolumn]{aastex631}

\newcommand{\fhd}{\ensuremath{f_\mathrm{HD}}}
\newcommand{\mbh}{\ensuremath{M_\mathrm{BH}}}

\usepackage{booktabs}

\submitjournal{ApJ}

\shorttitle{AGN and Stellar Contributions in Little Red Dots}
\shortauthors{Leung et al.}

\begin{document}

\title{Exploring the Nature of Little Red Dots: Constraints on AGN and Stellar Contributions from PRIMER MIRI Imaging}

\correspondingauthor{Gene C. K. Leung}
\email{gckleung@mit.edu}

\author[0000-0002-9393-6507]{Gene C. K. Leung}
\affiliation{Department of Astronomy, The University of Texas at Austin, Austin, TX 78712, USA}
\affiliation{MIT Kavli Institute for Astrophysics and Space Research, 77 Massachusetts Ave., Cambridge, MA 02139, USA}

\author[0000-0001-8519-1130]{Steven L. Finkelstein}
\affiliation{Department of Astronomy, The University of Texas at Austin, Austin, TX 78712, USA}

\author[0000-0003-4528-5639]{Pablo G. P\'erez-Gonz\'alez}
\affiliation{Centro de Astrobiolog\'{\i}a (CAB), CSIC-INTA, Ctra. de Ajalvir km 4, Torrej\'on de Ardoz, E-28850, Madrid, Spain}

\author[0000-0003-4965-0402]{Alexa M.\ Morales}
\altaffiliation{NSF Graduate Research Fellow}
\affiliation{Department of Astronomy, The University of Texas at Austin, Austin, TX 78712, USA}

\author[0000-0003-1282-7454]{Anthony J. Taylor}
\affiliation{Department of Astronomy, The University of Texas at Austin, Austin, TX 78712, USA}

\author[0000-0001-6813-875X]{Guillermo Barro}
\affiliation{Department of Physics, University of the Pacific, Stockton, CA 90340 USA}

\author[0000-0002-8360-3880]{Dale D. Kocevski}
\affiliation{Department of Physics and Astronomy, Colby College, Waterville, ME 04901, USA}

\author[0000-0003-3596-8794]{Hollis B. Akins}
\affiliation{Department of Astronomy, The University of Texas at Austin, Austin, TX 78712, USA}

\author[0000-0002-1482-5818]{Adam C. Carnall}
\affiliation{Institute for Astronomy, University of Edinburgh, Royal Observatory, Edinburgh, EH9 3HJ, UK}

\author[0000-0003-2332-5505]{\'{O}scar A. Ch\'{a}vez Ortiz}
\affiliation{Department of Astronomy, The University of Texas at Austin, Austin, TX 78712, USA}

\author[0000-0001-7151-009X]{Nikko J. Cleri}
\affiliation{Department of Astronomy and Astrophysics, The Pennsylvania State University, University Park, PA 16802, USA}
\affiliation{Institute for Computational and Data Sciences, The Pennsylvania State University, University Park, PA 16802, USA}
\affiliation{Institute for Gravitation and the Cosmos, The Pennsylvania State University, University Park, PA 16802, USA}

\author[0000-0002-3736-476X]{Fergus Cullen}
\affiliation{Institute for Astronomy, University of Edinburgh, Royal Observatory, Edinburgh, EH9 3HJ, UK}

\author[0000-0002-7622-0208]{Callum T. Donnan}
\affiliation{Institute for Astronomy, University of Edinburgh, Royal Observatory, Edinburgh, EH9 3HJ, UK}

\author[0000-0002-1404-5950]{James S. Dunlop}
\affiliation{Institute for Astronomy, University of Edinburgh, Royal Observatory, Edinburgh, EH9 3HJ, UK}

\author[0000-0001-7782-7071]{Richard S. Ellis}
\affiliation{Department of Physics \& Astronomy, University College London, London, WC1E 6BT, UK}

\author[0000-0001-9440-8872]{Norman A. Grogin}
\affiliation{Space Telescope Science Institute, 3700 San Martin Drive, Baltimore, MD 21218, USA}

\author[0000-0002-3301-3321]{Michaela Hirschmann}
\affiliation{Institute of Physics, Laboratory of Galaxy Evolution, Ecole Polytechnique Fédérale de Lausanne (EPFL), Observatoire de Sauverny, 1290 Versoix, Switzerland}

\author[0000-0002-6610-2048]{Anton M. Koekemoer}
\affiliation{Space Telescope Science Institute, 3700 San Martin Drive, Baltimore, MD 21218, USA}

\author[0000-0002-5588-9156]{Vasily Kokorev}
\affiliation{Department of Astronomy, The University of Texas at Austin, Austin, TX 78712, USA}

\author[0000-0003-1581-7825]{Ray A. Lucas}
\affiliation{Space Telescope Science Institute, 3700 San Martin Drive, Baltimore, MD 21218, USA}

\author[0000-0003-4368-3326]{Derek J. McLeod}
\affiliation{Institute for Astronomy, University of Edinburgh, Royal Observatory, Edinburgh, EH9 3HJ, UK}

\author[0000-0001-7503-8482]{Casey Papovich}
\affiliation{Department of Physics and Astronomy, Texas A\&M University, College Station, TX, 77843-4242 USA}
\affiliation{George P.\ and Cynthia Woods Mitchell Institute for Fundamental Physics and Astronomy, Texas A\&M University, College Station, TX, 77843-4242 USA}

\author[0000-0003-3466-035X]{{L. Y. Aaron} {Yung}}
\affiliation{Space Telescope Science Institute, 3700 San Martin Drive, Baltimore, MD 21218, USA}

\begin{abstract}

JWST has revealed a large population of compact, red galaxies at $z>4$ known as Little Red Dots (LRDs). We analyze the spectral energy distributions (SEDs) of 95 LRDs from the JWST PRIMER survey with complete photometric coverage from $1-18~\mu$m using NIRCam and MIRI imaging, representing the most extensive SED analysis on a large LRD sample with long-wavelength MIRI data. We examine SED models in which either galaxy or active galactic nucleus (AGN) emission dominates the rest-frame UV or optical continuum, extracting physical properties to explore each scenario’s implications. In the galaxy-only model, we find massive, dusty stellar populations alongside unobscured, low-mass components, hinting at inhomogeneous obscuration. The AGN-only model indicates dusty, luminous AGNs with low hot dust fractions compared to typical quasars. A hybrid AGN and galaxy model suggests low-mass, unobscured galaxies in the UV, with stellar mass estimates spanning $\sim$2 dex across the different models, underscoring the need for caution in interpreting LRD stellar masses. With MIRI photometry, the galaxy-only model produces stellar masses within cosmological limits, but extremely high stellar mass densities are inferred. The hybrid model infers highly overmassive black holes exceeding those in recently reported high-redshift AGNs, hinting at a partial AGN contribution to the rest-optical continuum or widespread super-Eddington accretion. Our findings highlight the extreme conditions required for both AGN or galaxy dominated scenarios in LRDs, supporting a mixed contribution to the red continuum, or novel scenarios to explain the observed emission.
\end{abstract}

\keywords{Active galactic nuclei (16) --- Galaxy formation (595) --- High-redshift galaxies (734) --- Supermassive black holes (1663)}

\section{Introduction} \label{sec:intro}

The commissioning of JWST \citep{gardner24} has opened a new window in the study of galaxies and supermassive black holes (SMBHs) in the first billion years of the universe. One of the most intriguing results from the first two years of JWST observations is the detection of a large population of red, compact objects at $z \gtrsim 4$ across multiple extragalactic surveys \citep[e.g.][]{labbe23a, akins23, furtak23, barro24, kocevski23, leung23}. These objects are characterized by their compact morphology and distinctive spectral energy distribution (SED), which features a red spectral slope in the rest-frame optical and a fainter blue continuum in the rest-frame UV, and are dubbed Little Red Dots \citep[LRDs, ][]{matthee24}.

Follow-up spectroscopy of LRD samples have revealed that a large fraction ($> 60\%$) display broad Balmer emission lines with velocity dispersions of $\sim 1000-2000\ \mathrm{km~s}^{-1}$ \citep{greene24, kocevski24, taylor24}, consistent with emission from the broad-line region (BLR) of active galactic nuclei (AGNs). In addition, the ALMA non-detection of dust emission at temperatures typical of star-forming galaxies further supports the AGN scenario \citep{lab23b}. The ubiquity of LRDs, $\sim 100$ times more common than UV-selected quasars at similar epochs \citep{greene24, akins24a, kokorev24a, kocevski24}, suggests that they could play a crucial role in the early growth history of SMBHs.

However, AGN activity is not the only interpretation of the physical nature of LRDs. While AGNs at up to $z \sim 6$ commonly produce an infrared excess at rest-frame $\sim 1-3 \mu$m due to the thermal emission the hot dusty torus in the vicinity of the SMBH \citep[e.g.][]{barvainis87, lyu22}, observations of some LRDs with JWST/MIRI have shown a flattening of the SED in these wavelengths \citep{williams23, pg23}, although some AGNs deficient of hot dust are known to exist from $z=0-6$ \citep{jiang10, hao10, lyu17}. In addition, the weakness in X-ray emission of LRDs in deep stacked Chandra imaging \citep{ananna24, yue24b} is unexpected for AGN with luminosities inferred from their rest-frame optical emission.

As an alternative, it has been suggested that an extremely compact and massive stellar distribution in LRDs can produce gas kinematics similar to those observed in the broad Balmer emission lines, and forbidden line emission can be suppressed in such dense environments, mimicking emission from the BLR of AGNs \citep{baggen24}. In fact, Balmer breaks, suggestive of an A-star-dominated rest-frame optical continuum, have been observed concurrent with broad Balmer emission lines in NIRSpec observations of a few LRDs \citep{wang24b, kokorev24b, ma24}. However, it has also been proposed that dense neutral gas around an AGN, as indicated by narrow absorption in the broad H$\alpha$ emission lines in some LRD spectra \citep[e.g.][]{matthee24, taylor24}, is sufficient to produce Balmer breaks in AGNs spectra that resemble those of stellar origin \citep{inayoshi24}. It remains unclear whether LRDs can be attributed to a unique population of AGN, massive compact galaxies, or a combination of both.

SED modeling can be a promising avenue to distinguish between the AGN and galaxy scenarios in LRDs by comparing model predictions of each case with the observed photometry \citep[e.g.][]{lab23b, pg23, barro24}. However, a simple ``goodness-of-fit'' test is inherently limited by the existing models of AGNs and galaxies used in the analysis---comparing deviation between the model and observation at face value is not straightforward, as significant systematic uncertainties in the models exist for novel objects such as LRDs.\footnote{This challenge is often depicted by the remark ``all models are wrong, but some are useful.'' \citep{box76}} Indeed, recent modeling of one very high signal-to-noise LRD continuum spectrum underscores this limitation \citep{ma24}. Rather than using SED modeling to accept or reject a model in its entirety, it can instead serve as a valuable tool to ``stress-test'' existing models by examining the implications under idealized assumptions. This approach can provide crucial constraints on the relative roles and contributions of AGN and galaxy scenarios in the physical nature of LRDs.

Long-wavelength MIRI data are critical in characterization of LRDs. While the distinctive two-component LRD SED in the rest-frame UV and optical is captured by the NIRCam bands, the long-wavelength filters of MIRI cover the rest-frame near-IR, which probes the hot dust associated with AGNs.
In this study, we perform the most extensive analysis of the rest-frame UV to near-IR SED of a sample of 95 LRDs using data from the JWST PRIMER survey \citep{dunlop21}, which uniquely provides NIRCam and MIRI imaging over $1-18\ \mu$m in a wide 200 sq. arcmin field. We test idealized SED models designed to represent extreme scenarios in which LRD emission is driven entirely by either AGN or galaxy light in the rest-frame UV and/or optical. Using these models, we derive galaxy and AGN properties of the LRDs, examine the implications of each scenario and constrain their respective contributions.

This paper is organized as follows. In Section \ref{sec:obs}, we describe the photometric and spectroscopic data used in our analysis. Section \ref{sec:sample} describes the selection and basic properties of the LRD sample. In Section \ref{sec:methods}, we describe our characterization of the LRDs by SED modeling and morphological analysis. We report the AGN and galaxy physical properties derived from the models in Section \ref{sec:results}. In Section \ref{sec:discussion}, we discuss the implications of the measured physical properties in the context of the AGN and galaxy population. Finally, we conclude and summarize our findings in Section \ref{sec:conclusions}.

Throughout this paper, we assume \citet{planck20} cosmology of $H_0=67.4\ \mathrm{km~s}^{-1}\ \mathrm{Mpc}^{-1}$, $\Omega_\mathrm{m}=0.315$ and $\Omega_\mathrm{\Lambda} = 0.685$. All magnitudes are in the AB system.

\section{Observations} \label{sec:obs}

\subsection{NIRCam Imaging}

The NIRCam imaging in the PRIMER UDS and COSMOS fields comes from the PRIMER survey, and is an internal reduction from the PRIMER team (internal version 0.6). Both fields contain the same set of NIRCam filters (F090W, F115W, F150W, F200W, F277W, F356W, F410M and F444W).  We also make use of HST/ACS F606W and F814W imaging in both fields.  

The photometry was measured following the process outlined in Finkelstein et al. (in prep.), focusing on measuring accurate colors and total flux estimates across HST/ACS, WFC3 and JWST/NIRCam imaging. Point spread functions (PSFs) are measured in each filters using stars in the stellar locus in the half-light radius versus magnitude space. Accurate colors are achieved by PSF-matching images with smaller PSFs than F277W to that band, and deriving correction factors for images with larger PSFs (via PSF-matching F277W to a given larger PSF).  Small Kron apertures are used to measure colors to optimize signal-to-noise for high-redshift galaxies.  Total fluxes are estimated by first deriving an aperture correction in the F277W band as the ratio between the flux in the larger (default) Kron aperture and the custom smaller aperture, with a residual aperture correction (typically $<$10\%) derived via source-injection simulations.  The key difference between the procedure described in  Finkelstein et al. (in prep.) and that used here is the inclusion of a ``hot$+$cold'' step, where first sources are selected with conservative (cold) detection parameters, designed to not split up large, bright galaxies.  Then, a more aggressive (hot) run is performed to identify fainter objects.  Objects from the hot catalog that fall outside the segmentation map from the cold run are added to the final photometry catalog. Photometric redshifts were estimated with \texttt{EAZY} \citep{brammer08}, using the same methodology and template set as described in Finkelstein et al. (in prep.), including the updated templates from \citet{larson23}. 

\subsection{MIRI Imaging}

The MIRI imaging used in this paper also comes from the PRIMER Team internal reductions (internal version 1.3.1). For the reduction of these data, we used the Rainbow JWST pipeline developed within the European Consortium MIRI GTO Team to deal with MIRI, NIRCam, and NIRISS imaging data, following the algorithms described in \citet[][see also \"Ostlin et al. 2024, submitted]{pg23}. Briefly, the method relies on the official JWST pipeline (version 1.11.4, 1130.pmap) but adds a superbackground strategy to homogenize the frames getting rid of horizontal and vertical stripes, as well as small-scale gradients. This important step has been shown to improve the depths of the MIRI images by several tenths of magnitude. In particular, for the PRIMER data, the superbackground method was configured so the 20 closest-in-time images were used to build the background model of a given single observation. This was an important tweak given that the PRIMER data were taken in several epochs (2 main per field) and significant changes in the background structure (especially in the bottom part of the field of view) and detector behavior were observed. A few pointings also showed an enhanced background level, linked to some outshining event that not only affected MIRI but also NIRCam (and resulted in aborted observations). For this reason, rather than using the data for the 2 epochs separately to build background frames, we constrained the superbackground datasets for a given image to the 20 closest observations (after some experimentation with number of images ranging from 10 to all the images in one epoch).

Another important tweak introduced in the Rainbow JWST pipeline to deal with PRIMER data consisted in calibrating the WCS of all the frames with the tweakreg package developed by the CEERS Team \citep{2023ApJ...946L..12B} and then switching this step off in the official pipeline stage 3 (mosaicking) execution. While the official pipeline tweakreg method introduced some systematic offsets between different frames at the 0.1\arcsec\, level, the CEERS version of the package was able to align all images more accurately and obtain a final mosaic with a typical WCS rms of 0.03\arcsec (i.e., one third of the MIRI pixel size).

The MIRI photometry for the little red dots was carried out as in \citet{pg23}. Circular aperture photometry was measured assuming several radii from 0.2 to 0.8\arcsec, applying aperture corrections for point-like sources. The flux of each source was obtained from the aperture presenting the highest S/N, being consistent with the photometry for smaller apertures within 1-$\sigma$, and whose aperture correction was smaller than 0.5~mag (typically achieved at radii 0.5\arcsec and 0.7\arcsec for F770W and F1800W, respectively). The background level and its rms were measured with the method described in \citet[][seel also \citealt{2008ApJ...675..234P}]{2023ApJ...951L...1P} to avoid the effects of correlated noise, building artificial apertures (with the same number of pixels used for the source) with non-contiguous pixels (for this work, pixels separated by more than 5 pixels, enough to avoid noise correlation introduced by the drizzling method used in mosaicking).

\subsection{NIRSpec Spectroscopy}

We supplement our photometric data with NIRSpec spectroscopy data from the RUBIES program \citep[JWST Cycle 2 GO\#4233, PIs de Graaff and Brammer;][]{degraaff24}. Specifically, we use the 9 currently available (as of August 2024) RUBIES NIRSpec pointings in the PRIMER-UDS field. All of these pointings were observed in the PRISM and G395M configurations, although not every targeted object was observed in both configurations. Each configuration and each pointing used three one-shutter nods and the NRSIRS2RAPID readout mode with 65 groups per nod (with a single integration per nod) for a total of 2889 seconds of exposure time.

We use the data reduction described in \cite{taylor24}, which we summarize below. \cite{taylor24} uses the JWST Calibration Pipeline (version 1.13.4 (DMS Build B10.1) and CRDS version 1215.pmap) in the default configuration (with the exception of a few \texttt{jump\_step} parameters custom tuned to better reject cosmic ray artifacts).

\begin{figure*}[thbp]
	\centering
		\includegraphics[width=\textwidth]{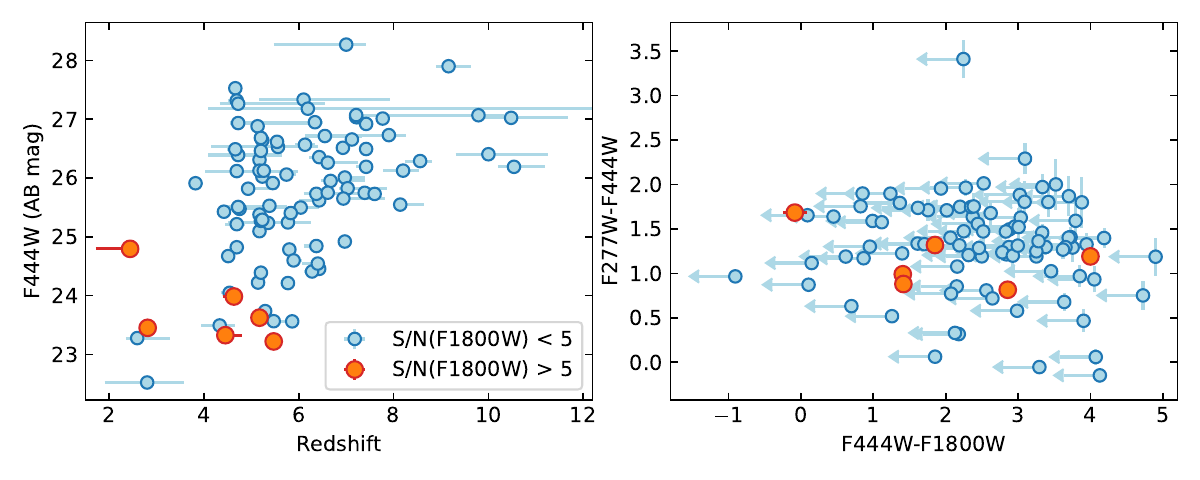}
		\caption{Left: The F444W magnitude versus photometric redshift for the LRD sample in this paper. The six LRDs with a significant detection in F1800W at S/N$>5$ are shown in the orange points. The horizontal error bars shows the 68\% confidence intervals of the photometric redshift. The LRDs detected in F1800W are generally bright, with F444W magnitude of $<25$ AB mag. Right: The F277W$-$F444W versus F444W$-$F1800W color-color diagram. The LRDs are red in F277W$-$F444W by selection. The LRDs detected in F1800W span a wide range of F444W$-$F1800W colors. For the rest of the sample, the $5\sigma$ upper limit on the F444W$-$F1800W are shown.}
		\label{fig:sample}
\end{figure*}

\section{Little Red Dot Sample} \label{sec:sample}

\subsection{Sample Selection}

We select our sample from a parent sample of LRDs identified in JWST legacy fields by \citet[][]{kocevski24}. We briefly describe the selection of the parent sample below, and we direct the readers to \citet{kocevski24} for the detailed selection methodology. 

\citet{kocevski24} presented a sample of 341 LRDs at $z\sim2-11$ selected by JWST/NIRCam imaging in the CEERS, JADES, NGDEEP, PRIMER and UNCOVER surveys. The LRDs were identified using continuum slope fitting on both sides of the Balmer break with shifting JWST and HST bandpasses by photometric redshift. This technique enables the detection of LRDs across a broader redshift range compared with previous studies that employed color selection criteria of fixed filter combinations. Sources with a red optical spectral slope of $\beta_\mathrm{opt} > 0$, a blue UV spectral slope of $\beta_\mathrm{UV} < -0.37$, and a signal-to-noise (S/N) of at least 12 in the F444W band were selected. As brown dwarfs are known to produce ``V-shape'' spectra similar to LRDs but with very blue UV colors, a lower limit is placed on the UV spectral slope $\beta_\mathrm{UV} > -2.8$ to remove contamination from brown dwarfs. Additionally, the LRDs were required to be compact in the F444W imaging, with a half-light radius from \texttt{SExtractor} of less than 1.5 times that of the stellar locus. To remove sources whose red optical spectral slopes likely resulted from strong emission lines, candidates were required to have $\beta_\mathrm{F277W-F356W} > -1$ and $\beta_\mathrm{F277W-F410M} > -1$.

From this parent sample, we select sources covered by MIRI imaging in the PRIMER survey in both the F770W and F1800W filters. This results in a sample of 101 LRDs, where 38 and 63 are located in PRIMER-COSMOS and PRIMER-UDS fields, respectively. One source has a close neighbor leading to contamination of its photometry, while five sources have part of their emission outside the edge of the MIRI image leading to inaccurate photometry. We remove these six sources from the sample. This leads to a final sample of 95 LRDs. 

We next cross-match our sample with the RUBIES PRIMER-UDS spectroscopic sample, and find 12 matches. We measure spectroscopic redshifts for this sub-sample by visually inspecting their spectra. We find nine objects with robust spectroscopic redshifts confirmed by multiple emission lines, but the remaining three objects show spectra that are either featureless or show only a single unidentified emission line. Throughout this work, we adopt the spectroscopic redshifts of these nine objects in place of their photometric redshifts. 

Our sample of LRDs spans $z \sim 2-11$, with the majority of the sources located between $z=4$ and 8. The LRDs cover a wide range in F444W magnitudes, from $\sim 22-28$ mag. We show the distribution of our sample in photometric (or spectroscopic, when available) redshift and F444W magnitude in the left panel of Figure \ref{fig:sample}.

\subsection{Mid-infrared Properties}

Within the LRDs in our sample, 83 are detected in the F770W filter with S/N$>5$.
In the F1800W MIRI filter, 17 sources are nominally detected with S/N$> 5$. However, the long wavelength filters of MIRI are known to be subject to non-Gaussian noise properties where individual pixels with high noise can skew the extracted photometry high. To remove spurious detections, we visually inspect the sources with S/N$>5$ in the F1800W filter. This results in the identification of 11 spurious sources, leaving 6 sources with a robust detection in the F1800W filter. For the spurious (S/N$>5$ but visually insignificant) and undetected (S/N$\le 5$) sources, we use the flux error to compute a $5\sigma$ upper limit to the F1800W flux.
In the right panel of Figure \ref{fig:sample}, we show the sample in a NIRCam-MIRI color-color diagram. All the LRDs display red colors of $\sim 0-2.5$ in F277W$-$F444W, which roughly corresponds to the rest-frame optical continuum slope, as expected from the sample selection.
For the six LRDs in our sample that are detected in the F1800W filter with S/N $>5$, the F444W$-$F1800W colors span $\sim 0-4$, with UDS 9235 showing the strongest emission in F1800W.
For the LRDs undetected in F1800W, the upper limits in F444W$-$F1800W span $\sim 0 - 4$, with the bluest color as low as $-1$. This flat mid-IR color is consistent with that observed in \citet{williams23}  and \citet{pg23}.

\begin{deluxetable*}{lcl}
%\vspace{2mm}
%\tabletypesize{\small}
\tablecaption{\textsc{Bagpipes} Priors for Galaxy-only Model}
\tablewidth{\textwidth}
\tablehead{\multicolumn{1}{l}{Parameter} & \multicolumn{1}{c}{Range} & \multicolumn{1}{c}{Description}}
\startdata
\multicolumn{3}{c}{\bf High-mass Component} \\
\texttt{sfh[``age'']} & (0.001, 15.0) & Age of stellar population in Gyr \\
\texttt{sfh[``tau'']} & (0.01, 10.0) & Delayed decay time in Gyr \\
\texttt{sfh[``massformed'']} & ($\log(M_0)\tablenotemark{a}-1$, 15.0) & $\log_{10}$ of the total stellar mass formed in $M_\odot$\\
\texttt{sfh[``metallicity'']} & (0.001, 2.0) & Matellicity in $Z_\odot$\\
\texttt{dust[``type'']} & \texttt{``Calzetti''} & Dust extinction law\\
\texttt{dust[``Av'']} & (0.0, 7.0) & Visual extinction in magnitude \\
\multicolumn{3}{c}{\bf Low-mass Component} \\
\texttt{sfh[``age'']} & (0.0001, 15.0) & Age of stellar population in Gyr \\
\texttt{sfh[``tau'']} & (0.01, 10.0) & Delayed decay time in Gyr \\
\texttt{sfh[``massformed'']} & (1.0, $\log(M_0)\tablenotemark{a} - 0.5$) & $\log_{10}$ of the total stellar mass formed in $M_\odot$\\
\texttt{sfh[``metallicity'']} & (0.001, 2.0), & Matellicity in $Z_\odot$\\
\texttt{dust[``type'']} & \texttt{``Calzetti''} & Dust extinction law\\
\texttt{dust[``Av'']} & (0.0, 2.0) & Visual extinction in magnitude\\
\enddata
\tablenotetext{a}{Preliminary mass to separate the high- and low-mass components. See Section \ref{sec:methods:gal} for details.}
%\tablecomments{comments.}
\label{tab:bagpipes}
%\vspace{-8mm}
\end{deluxetable*}

\section{Analysis} \label{sec:methods}

\subsection{Spectral Energy Distribution Modeling}

To explore the implications of the different origins of the LRD continuum emission, we model their SEDs with three different methods. These models are selected to represent the idealized scenarios of the possible origins of LRDs, where the emission is dominated by stellar or AGN light in the red and/or blue continua. First, we consider two models where the entire continuum from rest-frame UV through optical is dominated by \emph{only} stellar or AGN emission, respectively, which we call the galaxy-only model and AGN-only model. Additionally, we explore a hybrid model where the rest-frame optical is dominated by AGN emission and the rest-frame UV is dominated by stellar emission from the galaxy. For LRDs undetected in F1800W, we incorporate the $5\sigma$ upper limit on the flux in our SED fitting by adopting a piece-wise likelihood function that is flat below the $5\sigma$ upper limit, and traces the $5\sigma$ tail of a Gaussian above that.

\subsubsection{Galaxy-only Model}\label{sec:methods:gal}

For the galaxy-only model, we use a modified version of \textsc{Bagpipes} \citep{car18}, a Bayesian code for modeling and fitting galaxy photometry and spectra. Modifications have been made to allow for the simultaneous fitting of two star formation histories with distinct dust extinction parameters. We use this to model two stellar populations, a high-mass component that dominates the emission in the rest-optical and a potential, second low-mass component that dominates the rest-UV continua. We include two stellar populations, each with a separate delayed-tau star formation history and a separate \citet{cal00} dust attenuation law. To separate the high-mass and low-mass components, we set different boundaries for the stellar mass formed for each component.
The stellar mass boundary for the two components is determined as follows. We first perform an initial round of SED fitting to the rest-optical photometric filters only and measure the stellar mass formed for this initial fit. This preliminary mass serves as a prior estimate of the stellar mass formed of the high-mass component. In the main fitting round, the stellar mass formed of the high-mass component is required to be no less than one dex below this preliminary mass, while that of the low-mass component is set to have a maximum of 0.5 dex below the preliminary mass. We report all the priors used in Table \ref{tab:bagpipes}.

\subsubsection{AGN-only Model}

\begin{figure}[!t]
	\centering	\includegraphics[width=0.5\textwidth]{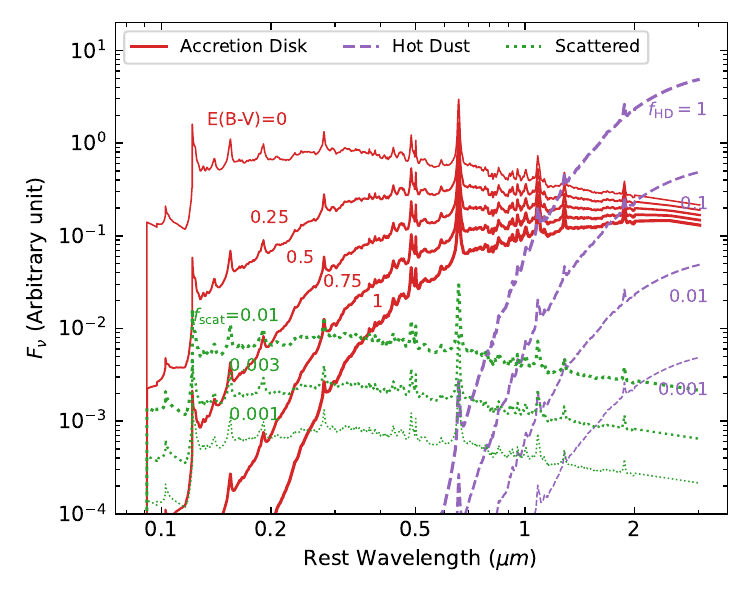}
		\caption{Visualization of the components in our AGN-only SED model based on the quasar templates in \citet{tem21}. The red solid lines show the accretion disk emission subject to a varying degree of dust reddening. The violet dashed lines show the emission from the hot dust at 1240 K with variable hot dust fractions (\fhd ). The green dotted lines show the scattered light represented by a fraction of the unobscured accretion disk emission.}
		\label{fig:mod}
\end{figure}

\begin{figure*}[thbp]
	\centering
		\includegraphics[width=\textwidth]{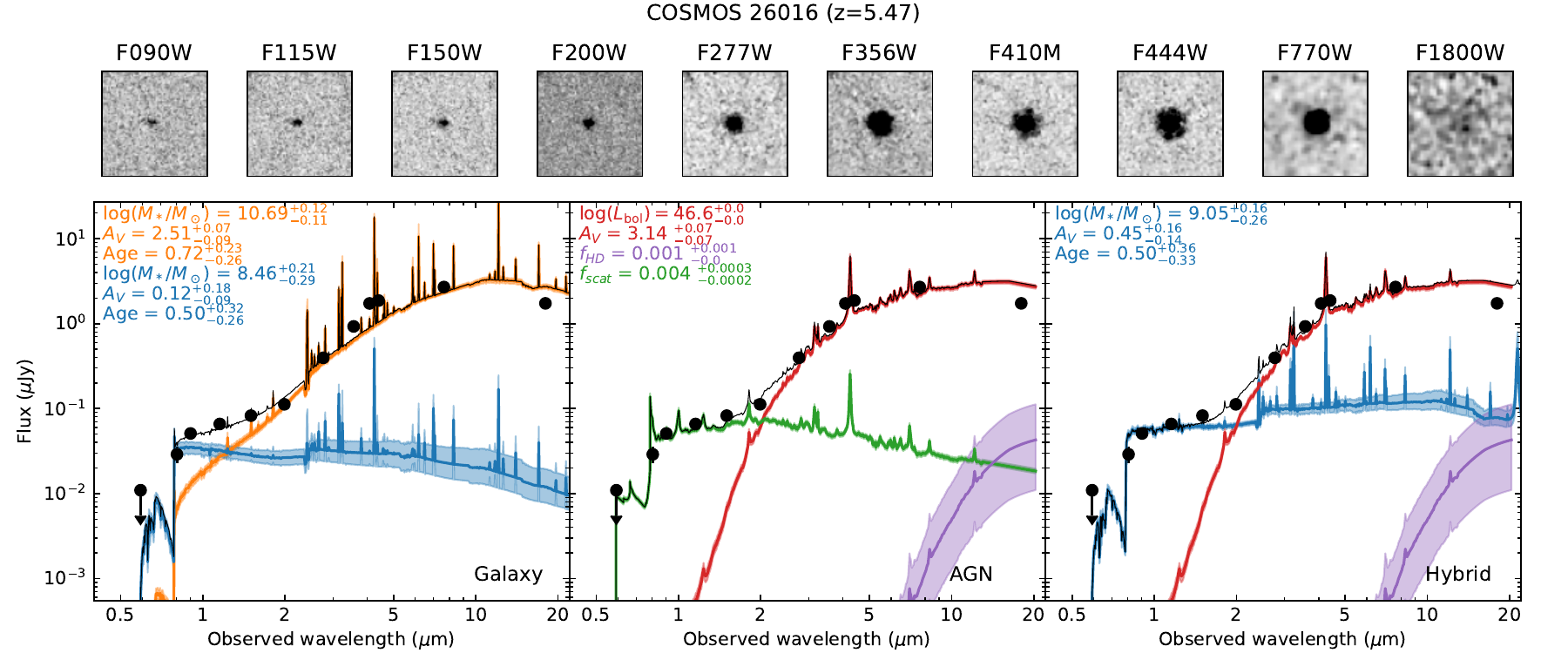}
		\includegraphics[width=\textwidth]{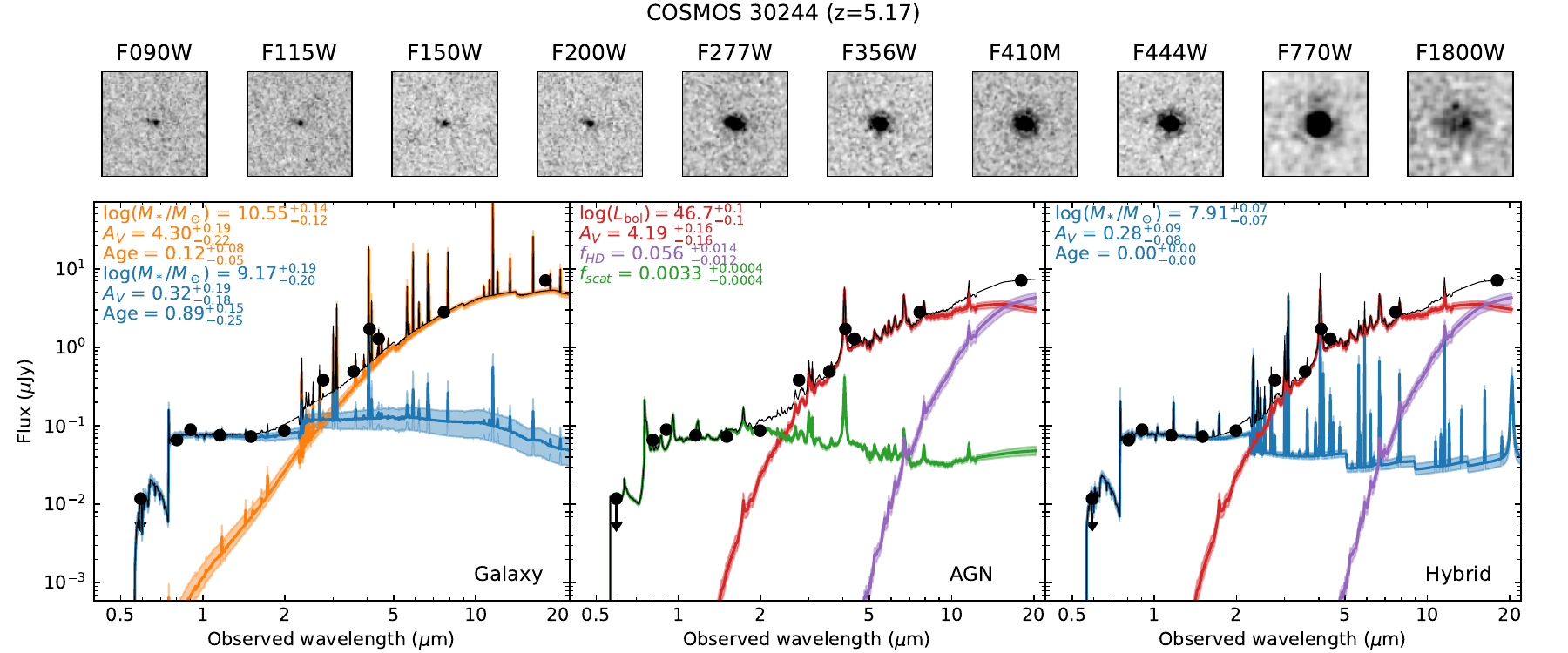}
		\caption{Image cutout and SED of the six LRDs detected in F1800W in our sample. The top panel shows 2'' stamp images. The bottom panel shows the observed photometry in the black points, and $2\sigma$ upper limits for non-detections, except for F1800W, where $5\sigma$ upper limits are shown for non-detections. The median (colored lines) and 68\% posterior (shaded regions) of the galaxy-only, AGN-only and hybrid models are shown from left to right. The rest-UV stellar component is shown in blue, rest-optical stellar component in orange, reddened accretion disk emission in red, scattered accretion disk in green, and hot dust emission in violet. The complete figure set (95 images) is available in the online journal.}
		\label{fig:sed}
\end{figure*}

\begin{figure*}[thbp]
    \figurenum{\ref{fig:sed}}
	\centering
		\includegraphics[width=\textwidth]{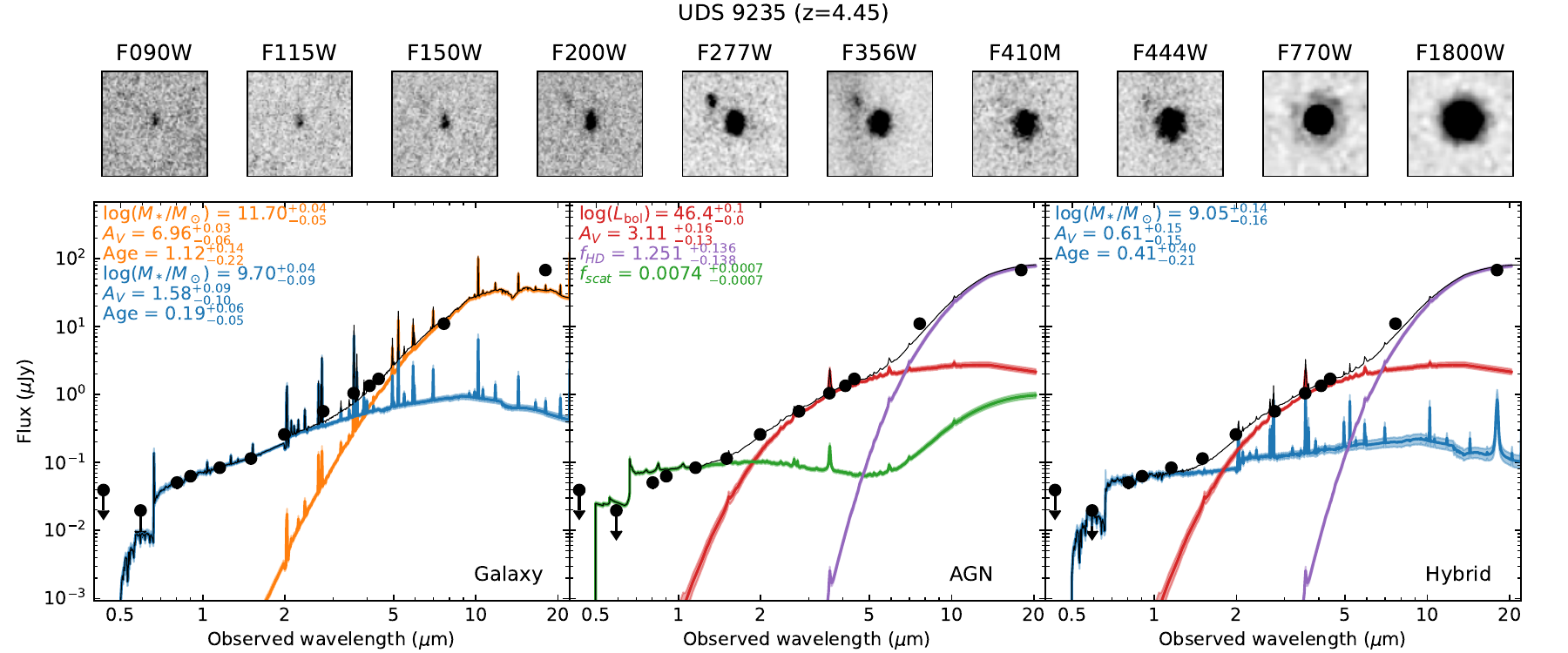}
		\includegraphics[width=\textwidth]{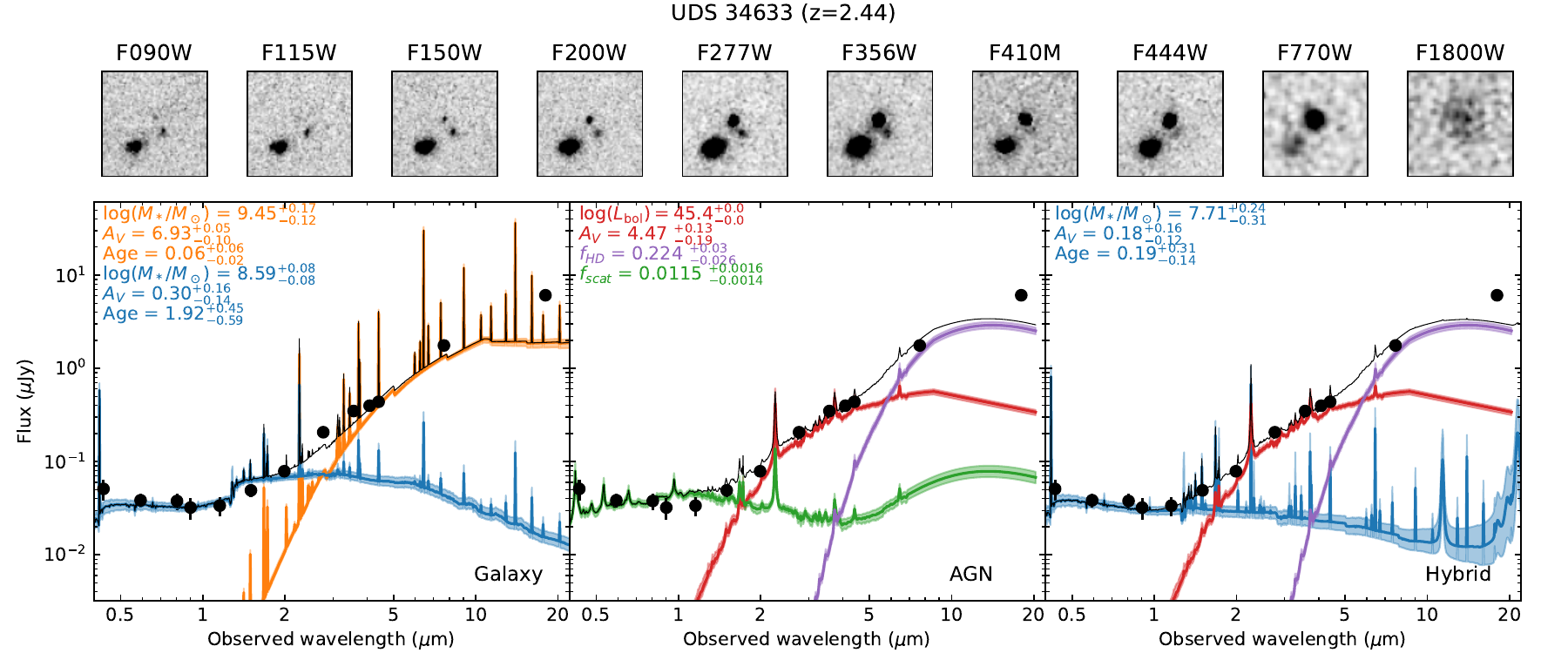}
		\caption{Continued.}
\end{figure*}

\begin{figure*}[thbp]
    \figurenum{\ref{fig:sed}}
	\centering
		\includegraphics[width=\textwidth]{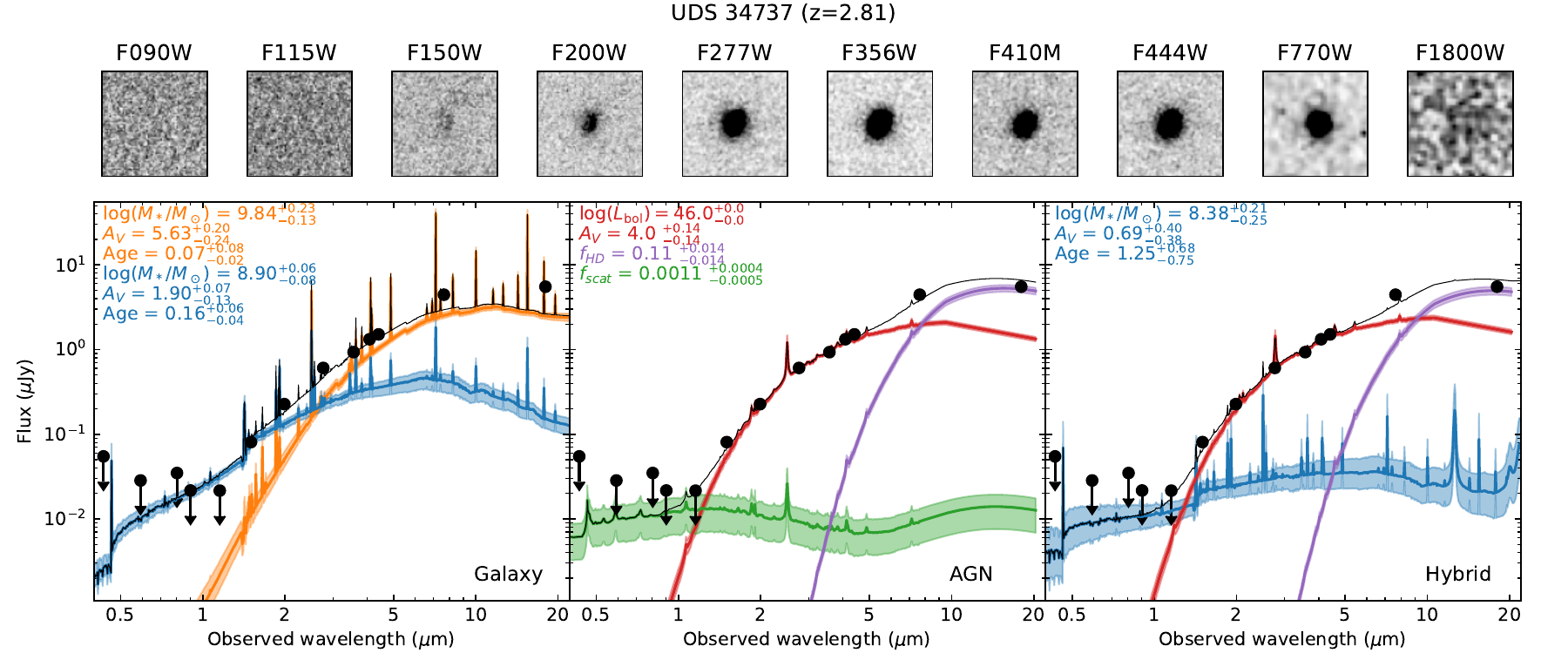}
		\includegraphics[width=\textwidth]{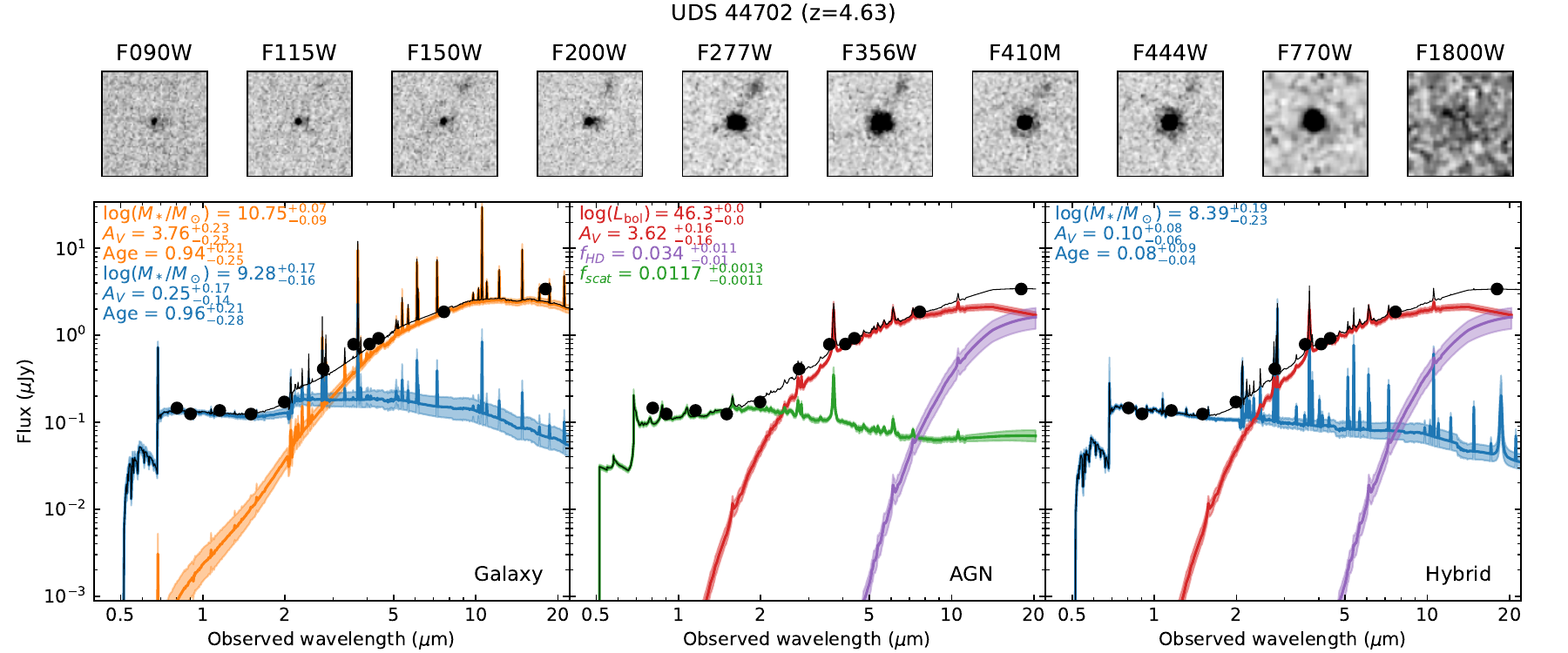}
		\caption{Continued.}
\end{figure*}

For the AGN-only model, we assume a reddened AGN with scattered light picture similar to the physical scenario presented  \citet{lab23b} and \citet{greene24}. In this picture, the rest-optical to infrared continuum originates from a dust-reddened accretion disk and/or the hot torus dust, while the rest-UV is produced by scattered emission of the intrinsic, unobscured accretion disk. We model the SED using the type 1 quasar templates in \citet{tem21} produced by the \texttt{qsogen} code. 

In these templates, the accretion disk is modeled with a broken power law, where the indices are fixed to the best-fit values from the sample of $z=0-5$ quasars in \citet{tem21} and the normalization is tied to the monochromatic luminosity at rest-frame $3000$ \AA\ prior to dust extinction ($L_{3000}$). We also include the average quasar emission line template in \citet{tem21} (\texttt{emline\_type}=0), allowing the equivalent widths (EWs) of the lines to vary by a scaling factor $f_\mathrm{eml}$. 
The accretion disk and emission lines are subject to a common dust attenuation based on the quasar dust attenuation law from \citet{tem21} with a variable $A_V$, which is similar to the SMC extinction curve, except having a shallower slope at $\lesssim 1700$ \AA. 

In addition, a variable hot dust component is included, modeled as a single black body to represent the emission from the AGN torus. The temperature of the hot dust is fixed at $T=1240$ K, the best-fit value found in $z=0-5$ quasars in \citet{tem21} and similar to the typical sublimation temperature of granite and silicate dust grains \citep[e.g.][]{barvainis87}. The strength of the hot dust component is controlled by a hot dust fraction parameter ($f_\mathrm{HD}$), defined as the peak of the black body spectrum relative to the average of the quasar sample in \citet{tem21}. To allow for different sources of dust obscuration, the hot dust component and the dust attenuation parameters are independent of each other, so we do not assume that the reddening is produced by the torus dust. Finally, we include a scattered component, representing the intrinsic accretion disk spectrum escaping through the obscuring medium unattenuated. This component is controlled by a scatter fraction parameter ($f_\mathrm{scat}$), defined as the normalization of the scattered component relative to the unobscured accretion disk. This results in five free parameters, namely $L_{3000}$, $f_\mathrm{eml}$, $A_V$, $f_\mathrm{HD}$ and $f_\mathrm{scat}$. We use the Markov chain Monte Carlo implementation \texttt{emcee} to sample the free parameters. We visualize the various components of the model in Figure \ref{fig:mod}.

\subsubsection{Hybrid Model}

Finally, we consider a hybrid model with mixed stellar and AGN contributions. In this model, we assume the rest-optical continuum is dominated by a reddened AGN, while the rest-UV light is produced by the stellar emission of the galaxy. This is motivated by the results of \citet{kil23} and \citet{kocevski24}, which found an unresolved morphology in the rest-optical and extended emission in the rest-UV. \citet{kil23} also found a high Balmer decrement in the broad emission lines compared to the narrow lines, suggesting a compact, obscured origin for the red continuum and an extended, unobscured source for the blue. We note that we do not attempt to decompose potential mixed AGN and galaxy in the rest-optical continuum for each individual object in this exercise, but we instead use the idealized assumption of maximum AGN contribution in this wavelength regime to examine the implications of this picture, providing constraints on its contribution at the population level.

For the hybrid model, we adopt the AGN-only model fit from the previous section, removing the scattered component in the rest-UV, before subtracting the median model fluxes from the observed photometry to obtain the residual in the rest-UV. We convolve the uncertainty of the residual photometry with the $68\%$ uncertainties of the model fluxes. We then fit the residual photometry using a single stellar population with a delayed tau star formation history and a \citet{cal00} dust attenuation law with \textsc{Bagpipes}. The final model thus includes the red component from the AGN-only model in the rest-optical and a blue stellar-only model in the rest-UV.

\begin{figure*}[thbp]
	\centering
		\includegraphics[width=\textwidth]{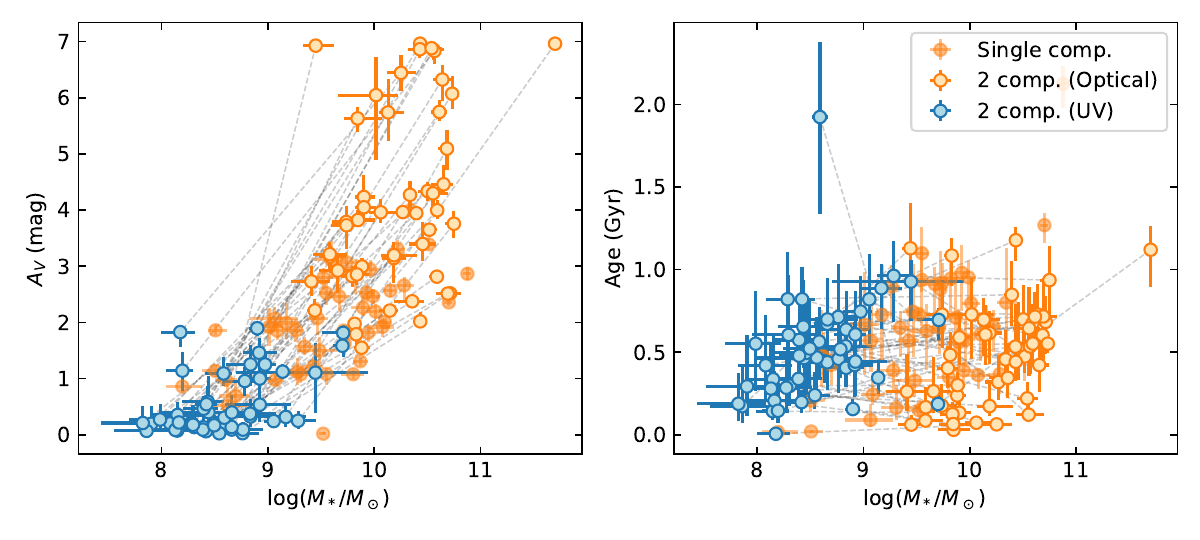}
		\caption{Galaxy properties from the two-component galaxy-only fit. For LRDs where two components are significantly detected, the rest-UV component is shown in blue, and the rest-optical component is shown in orange. The two components of the same LRD are connected by grey dotted lines. LRDs where only one component is required are shown in the solid orange points. Left: Distribution of $A_V$ versus stellar mass. Right: Distribution of age versus stellar mass. The rest-UV component are substantially less dusty than the rest-optical component, but have similar ages as the latter. This is consistent with the picture that the rest-UV continuum originates from the scattered light of the same stellar population that produces the rest-optical emission.}
		\label{fig:gal}
\end{figure*}

\subsection{Morphological Analysis}

To measure the stellar mass density of the LRDs, we use \textsc{galfit} \citep{peng2002} to model the galaxy light profiles in the F444W filter. For each galaxy, we provide the measured total magnitude, 1.0\arcsec\ F444W image stamps centered on the source, along with the corresponding error images, 1.0\arcsec\ segmentation maps to mask light from nearby sources, and the F444W PSF. Each object is fitted with two models: a PSF profile and a Sérsic profile convolved with the PSF (See \citealt{peng2002} for galaxy parameters fit for each profile). \textsc{galfit} returns the best-fit galaxy parameters, the corresponding least-$\chi^2$ value, a model image stamp for the source, and a residual map showing the difference between the data and the model. \textsc{galfit} also allows us to impose constraints on specific parameters and provide initial guesses to guide the fitting process. In this work, we constrain the following: (1) $x$ and $y$ position, (2) integrated magnitude, $23 \leq \mathrm{mag} \leq 30$, (3) effective radius, $0.03 \leq R_e \leq 5.0$ pixels, (4) axis ratio, $0.25 \leq b/a \leq 20$, (5) position angle, $0^\circ \leq \mathrm{PA} \leq 90^\circ$, and (6) for the Sérsic profile only, Sérsic index, $0.5 \leq n \leq 10$.
We adopt the Sérsic fit if the difference in the Bayesian Information Criterion (BIC) is greater than 6, suggesting strong evidence favoring the Sérsic model, and if the relative uncertainty in $R_e$ is less than unity. Using these criteria, the Sérsic model is preferred in three sources, while the PSF fit is sufficient for the remaining 92 sources.

\section{Results} \label{sec:results}

In this section, we discuss the galaxy and AGN properties derived from our SED fitting. Figure \ref{fig:sed} shows the cutouts and SEDs of the six LRDs detected in F1800W. The complete figure set for the full sample is available in the online journal.

\begin{figure*}[thbp]
	\centering
    \includegraphics[width=\textwidth]{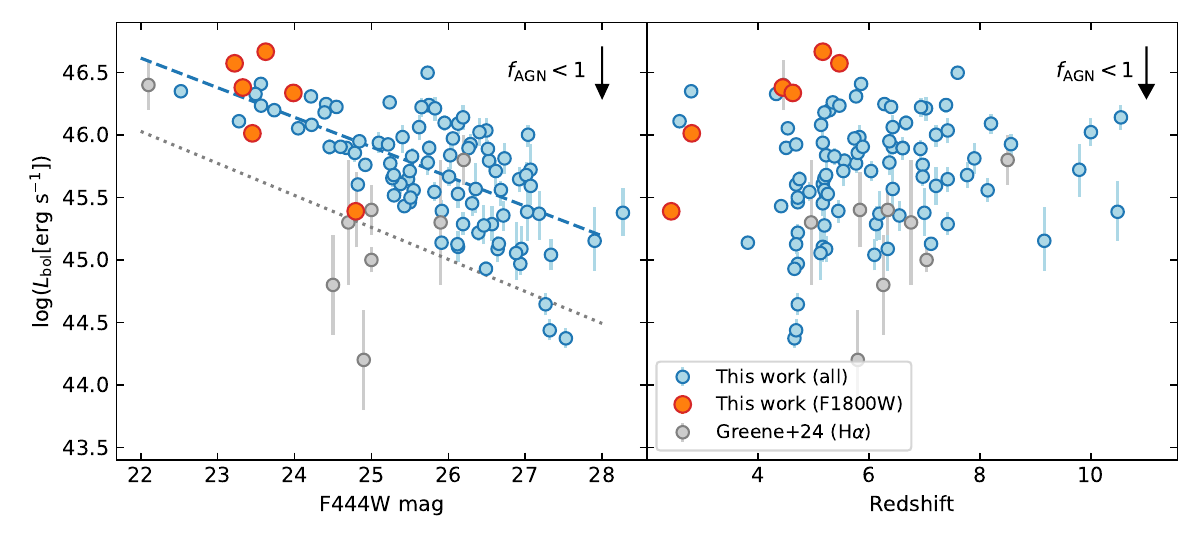}
		\caption{AGN bolometric luminosity in the AGN-only fit versus F444W magnitude (left) and photometric redshift (right). The LRDs detected in F1800W are shown in the orange points. AGN bolometric luminosity of the LRDs measured by the H$\alpha$ emission line in \citet{greene24} are shown in the grey points. The blue dashed (grey dotted) line shows the best-fit linear relation between the AGN bolometric luminosity and F444W magnitude to the data in this study \citep[in][]{greene24}. The downward arrows indicate the direction in which the bolometric luminosity would change if the Eddington ratio is decreased. The derived AGN bolometric luminosity in this study follows a very similar slope as that in \citet{greene24}, but is systematically higher than  by 0.6 dex. This could suggest that the AGN contribution to the rest-optical continuum is less than unity.}
		\label{fig:lbol}
\end{figure*}

\subsection{Galaxy-only Fit}

In the two-component galaxy-only fit, the high-mass component, which dominates the rest-frame optical, accounts for the majority of the stellar mass of the systems, as it dominates the overall bolometric output of the LRD. We define the potential second low-mass component in the rest-frame UV as significant when the $68\%$ credible interval of the stellar mass of this component is smaller than two dex. Otherwise, we consider that only one component is required.

In Figure \ref{fig:gal}, we show the distribution of the dust extinction and ages from the SED modeling as a function of stellar mass for the galaxy-only fit. For 45 of the 95 LRDs, only the high-mass component is required. For these sources, we find moderate to high $A_V$ of $1-4$, while the ages show a wide distribution spanning $0.002-1$ Gyrs. For the remaining 50 LRDs both the high- and low-mass components are detected. For these sources, we find a low-dust, low-mass component in combination with a dusty, high-mass component.
The low-mass component has a lower stellar mass than that of the high-mass component by a median 1.6 dex. The low-mass components also show significantly lower dust extinction of $A_V \sim 0-1$ compared with $A_V \sim 2-4$ for the high-mass components. For a small number of sources, the high-mass component exhibit extreme inferred dust extinction of $A_V > 6$. Interestingly, when two components are required, the ages of the two components are not significantly different from one another, although the uncertainty is high. The different dust attenuation but similar age is consistent with the picture that the rest-UV continuum comes from the scattered light of the same stellar population that produces the rest-frame optical emission in patchy obscuration.

\subsection{AGN-only Fit}\label{sec:res:agn}

\begin{figure*}[thbp]
	\centering
		\includegraphics[width=\textwidth]{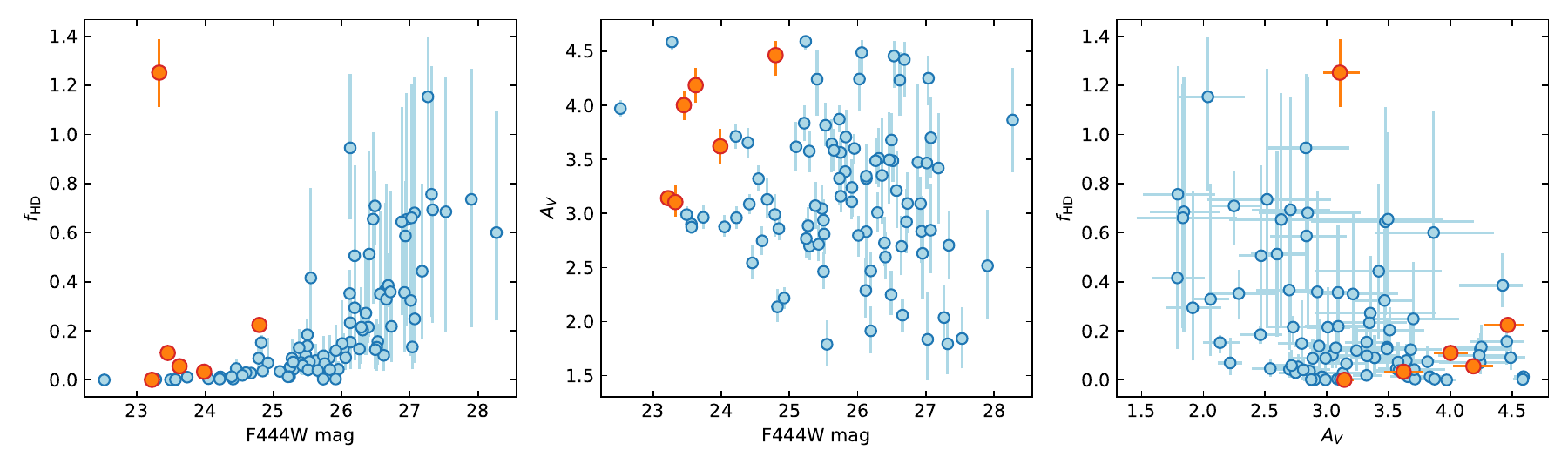}
		\caption{AGN properties derived from the AGN-only SED fits. Left: The hot dust fraction versus F444W magnitude. Middle: $A_V$ vs F444W magnitude. Right: Hot dust fraction versus $A_V$. The hot dust fraction is low ($\lesssim 0.2$) in the majority of the LRDs, with the exception of UDS 9235. Our results also show that high $A_V$ of  $\gtrsim 2$, consistent with a dust obscured AGN producing the rest-optical continuum. There is no correlation between \fhd\ and $A_V$, suggesting that the dust responsible for the reddening of the AGN does not reside in a hot dust torus.}
		\label{fig:agn}
\end{figure*}
In the AGN-only fit, the model parameters include the intrinsic monochromatic luminosity at 3000 \AA\ prior to dust extinction $L_{3000}$, the dust extinction measured by $A_V$, the strength of hot dust emission measured by \fhd , the amount of scattered light contributing to the rest-frame UV continuum as $f_\mathrm{scat}$, as well as the emission line scaling factor. We convert $L_{3000}$ to the intrinsic AGN bolometric luminosity using the bolometric corrections in \citet{run12, run12b}. In Figure \ref{fig:lbol}, we show the AGN bolometric luminosity of the LRDs as a function of the F444W magnitude and redshift. The intrinsic AGN bolometric luminosity spans $\sim 10^{45-46}~\mathrm{erg~s}^{-1}$. As expected by the design of the SED model, the bolometric luminosity is strongly driven by the F444W magnitude. 

For comparison, we also show the bolometric luminosity derived from H$\alpha$ for the spectroscopic sample of LRDs in \citet{greene24} and the broad-line AGNs in \citet{matthee24}. We also plot the best-fit linear relation between the bolometric luminosity and F444W magnitude for our sample and that of \citet{greene24}. While our SED-derived bolometric luminosity and that of the spectroscopic samples follow a similar trend with F444W, our values are systematically higher than the latter by $\sim 0.6$ dex. This could suggest that the SED-derived bolometric luminosity is overestimated. Since our AGN-only SED model assumes that the entirety of the rest-optical continuum is produced by the AGN, a lower bolometric luminosity in line with the spectroscopic results can be explained by a lower contribution of the AGN to the red continuum, implying that at least some of the red continuum may be produced by stellar emission. The bolometric luminosity does not show any significant trends with redshift.

\begin{figure*}[thbp]
	\centering
    \includegraphics[width=0.95\textwidth]{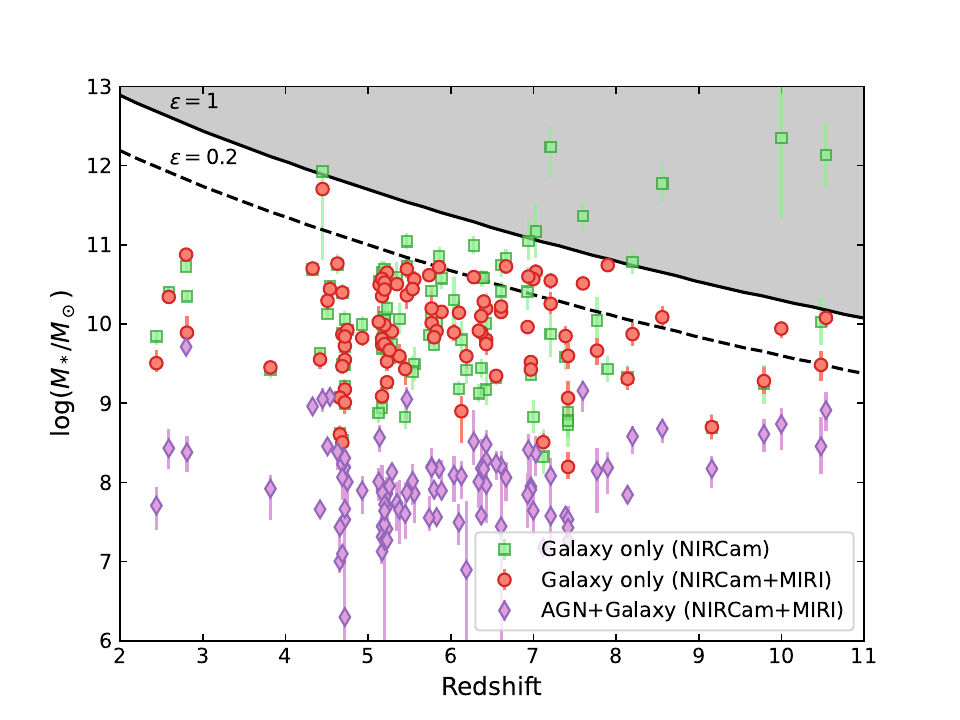}
		\caption{Stellar mass versus redshift for the LRDs in this study. The stellar masses derived from the galaxy-only fit without MIRI photometry are shown in green, galaxy-only fit with MIRI photometry in red, and hybrid fit (where the rest-optical emission is attributed to an AGN) in violet. We also include the limiting stellar mass expected for baryon conversion efficiencies of 1 and 0.2 in the black solid and dashed lines, respectively. The inclusion of MIRI photometry in the galaxy-only fit significantly reduces the stellar mass of LRDs at $z \gtrsim 7$, bringing them back to the expected range. Very high baryon conversion efficiencies of $>0.3$ are still inferred in 10 LRDs when considering the galaxy-only model with MIRI photometry, suggesting that the galaxy-only model could overestimate their stellar mass. With the hybrid model, the stellar mass of all LRDs are within expectations.}
		\label{fig:mstar}
\end{figure*}

\begin{figure*}[!t]
	\centering
    \includegraphics[width=0.75\textwidth]{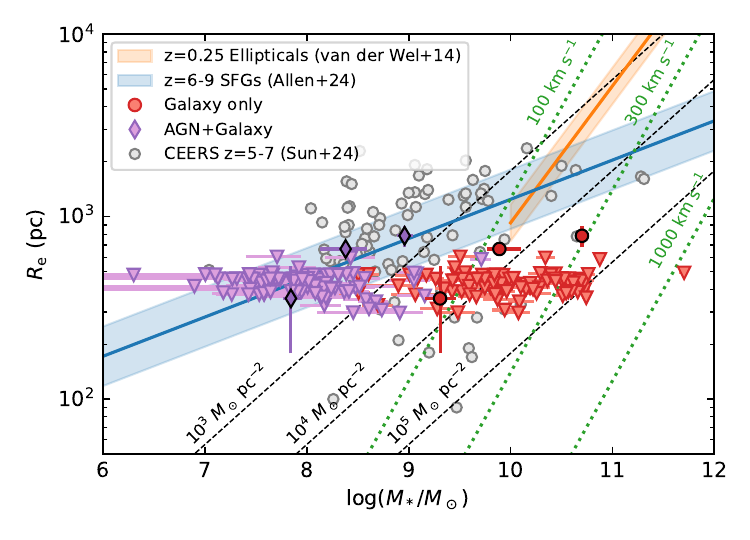}
		\caption{Effective radius in the F444W filter versus stellar mass. The red symbols show the stellar mass estimates from the galaxy-only fit, while the violet symbols show those from the hybrid fit. Upper limits using the HWHM of the F444W filter are shown with triangles for the 92 LRDs that are not significantly resolved. The three resolved LRDs are highlighted by black outlines. Values for a collection of $z=5-7$ galaxies in CEERS from \citet{sun24} are shown in the grey points. The best-fit size-mass relations for elliptical galaxies at $z=0.25$ \citep{vanderwel14} and typical star-forming galaxies at $z=6-9$ \citep{allen24}, and their $1\sigma$ scatter are shown in the orange and blue lines and shaded regions, respectively. We also show lines of constant predicted velocity dispersion from \citet{baggen24} in the green dotted lines. The stellar mass estimates from the galaxy-only model results in size-mass relations deviating from the $z=6-9$ best-fit by $>1-2\sigma$, implying high stellar mass densities of $\sim 10^5\ M_\odot~\mathrm{pc}^{-2}$. These high densities are also $1-2$ dex higher than local ellipticals, suggesting a drastic evolution in stellar mass distribution is required to produce present-day galaxies. The hybrid fit results in lower stellar masses that are largely in agreement with the best-fit relation at $z=6-9$ within its intrinsic scatter. Stellar mass densities inferred from the galaxy-only model are capable to producing velocity dispersions of at least a few hundred km~s$^{-1}$, similar to broad emission lines in LRDs.}
		\label{fig:re}
\end{figure*}

In Figure \ref{fig:agn}, we show the distribution of the \fhd\ and $A_V$ from our SED modeling results. The vast majority of the sources show very low hot dust fractions of $\lesssim 0.2$, with the exception of UDS 9235, which has $\fhd \sim 1$. This shows that the typical emission signature from the hot dust torus at $T \sim 1200$ K is absent in the LRDs. For sources fainter than $\sim 26$ mag in F444W, the hot dust fraction is largely unconstrained due to the low brightness of the source. The LRDs also display strong dust extinction of $A_V \sim 2-4.5$, consistent with an obscured accretion disk producing the observed red optical continuum. We do not observe any correlation between the hot dust fraction and dust extinction. This implies that the dust responsible for the reddening of the AGN does not reside in a hot dust torus. Finally, we note that sources with or without an F1800W detection occupy a similar parameter space in \fhd\ and $A_V$ (with the exception of UDS 9235). This suggests that the detection of F1800W in those sources are primarily due to their overall brightness and not their AGN properties.

\subsection{Hybrid Fit}

In the hybrid fit, we explore the scenario where the rest-optical continuum originates from an obscured AGN, and the rest-UV is produced by stellar emission. We find that this model infers a low-mass, low dust star forming galaxy responsible for the rest-UV emission. The derived stellar masses are lower than the total stellar masses in the galaxy-only model by a median of 1.9 dex. The dust extinction is generally low, with $A_V \lesssim 0.5$, and the stellar populations have young ages at a median of $\lesssim 0.2$ Gyrs.

\subsection{UDS 9235}

An outlier in our sample is UDS 9235. It displays significant rest-frame near-IR excess, with a F1800W$-$F444W color of $\sim 4$ which is the reddest within the sample. With a measured hot dust fraction of $\fhd = 1.3 \pm 0.1$, it has similar hot dust emission properties to average AGNs at $z=0-5$, suggesting that it likely represents a typical obscured AGN. It has a UV spectral slope of $\beta_\mathrm{UV} = -0.73 \pm 0.35$, which is redder than the median $\beta_\mathrm{UV}$ of $-1.77$ in the LRD sample selected by \citet{kocevski24}, but is within the selection criterion of $\beta_\mathrm{UV} < -0.37$. It has a F115W$-$F150W color of 0.34, which is also within the color selection criterion of $\mathrm{F115W}-\mathrm{F150W}<0.8$ in \citet{greene24}. The markedly different observed mid-IR properties of UDS 9235 to the rest the sample suggests that it could be of distinct physical origins to the general LRD population. This highlights that photometric selection of LRDs can potentially yield a disparate population of objects \citep{hainline24}.

\section{Discussion} \label{sec:discussion}

A crucial question regarding the nature of LRDs is the origin of the rest-optical continuum, which gives rise to their distinctive SED shape and dominates their energy output. In this section, we consider the two possible interpretations of LRDs as originating dominantly from stellar or AGN emission. We discuss of implications of both scenarios to the evolution of galaxies and SMBHs.

\subsection{Stellar Interpretation}

In Figure \ref{fig:mstar}, we show the stellar mass estimates of the LRDs using the galaxy-only and hybrid models as a function of redshift. In addition, to investigate the impact of the inclusion of MIRI photometry on stellar mass estimates, we also show results from a separate galaxy-only fit using only NIRCam and HST photometry. To place the results in context, we also show the theoretical stellar mass limit based on the evolution of the halo mass function. We estimate this by integrating the halo mass function to estimate the mass above which one halo is expected from our survey volume assuming sliding redshift bins of $\Delta z = 2$. We use the halo mass function calculator \texttt{hmf} \citep{mur13, mur14} and the halo mass function of \citet{beh13}, and assume a cosmic baryon fraction of 0.158. We show two stellar mass limits with a baryon conversion efficiency ($\epsilon$) of 1 and 0.2, respectively. 

For the galaxy-only model, the inclusion of MIRI photometry significantly reduces the stellar mass estimate for LRDs at at $z \gtrsim 7$, where the Balmer break begins to be redshifted out of the NIRCam coverage. The stellar mass of LRDs at these redshifts decreases by a median of 0.38 dex when MIRI photometry is taken into account. A similar decrease is found in galaxies at $z=6-9$ in the CEERS survey \citep{papovich23}. Without including MIRI photometry, nine of the LRDs are found to be improbably massive, exceeding the theoretical stellar mass limit with $\epsilon = 1$. By contrast, with the inclusion of MIRI photometry, no LRDs exceed this limit. Nonetheless, even with the inclusion of MIRI photometry, very high baryon conversion efficiencies of $\epsilon > 0.2$ are implied in 14 of the LRDs assuming the galaxy-only model, while four LRDs require extreme efficiencies of $\epsilon > 0.5$. This shows that the galaxy-only model could overestimate the stellar mass even when considering the MIRI photometry, though there is complementary evidence that the baryon conversion efficiency is higher at $z > 4$ \citep{chworowsky24}. In this case, LRDs could represent a plausible progenitor to massive quiescent galaxies discovered at $z\sim 4$ \citep[e.g.][]{carnall24, degraaff24b}. When considering the hybrid model, the stellar mass inferred is $\sim 2$ dex lower than that of the galaxy-only model with MIRI photometry, with all LRDs falling under the 20\% baryon conversion efficiency limit. Our results show that LRDs do not pose a challenge to mass assembly under our current cosmological model when MIRI photometry is considered. When MIRI photometry is included, the stellar mass estimates are highly model-dependent, subject to a $\sim 2$ dex uncertainty. 

In Figure \ref{fig:re}, we plot the effective radius in the F444W filter as a function of stellar mass for the galaxy-only and hybrid models. For the 92 sources which are not significantly resolved, we show upper limits of the effective radius derived from the HWHM of the PSF of the F444W filter and the redshift of the source. For the unresolved sources, the galaxy-only model implies upper limits of the stellar mass surface density ($\Sigma_*$) with a median of $\sim 2\times 10^4~M_\odot\ \mathrm{pc}^{-2}$, while the hybrid model returns a median upper limit of $\sim 200~M_\odot\ \mathrm{pc}^{-2}$. For the three resolved sources, the galaxy-only and hybrid models return $\Sigma_*$ of $0.5-2.6\times 10^4~M_\odot\ \mathrm{pc}^{-2}$ and $200-500~M_\odot\ \mathrm{pc}^{-2}$, respectively. For context, we also plot the best-fit size-mass relation for galaxies at $z=6-9$ in \citet{allen24} and its $1\sigma$ intrinsic scatter, as well as those for elliptical galaxies at $z=0.25$ in \citep{vanderwel14}. With the galaxy-only model, 92 of the LRDs have size upper limits that are $1-4\sigma$ away from the best-fit $z=6-9$ relation compared with the intrinsic scatter, implying extremely high densities relative to typical galaxies. When using the hybrid model, the size upper limits of $80\%$ of the LRDs lie within the $1\sigma$ intrinsic scatter of the size-mass relation. This again highlights the disparate conclusions resulting from the $\sim 2$ dex uncertainty in stellar mass depending on the model employed. When compared with local elliptical galaxies, the majority of LRDs have stellar mass densities that are at least $1-2$ dex higher. This suggests that a drastic evolution in stellar mass density is required to reach the properties observed in present-day galaxies. We also show in Figure \ref{fig:re} lines of constant velocity dispersion from \citet{baggen24}. We find that the galaxy-only model produces stellar mass density lower limits that can generate velocity dispersions of at least several hundred km~s$^{-1}$, consistent with those observed in the broad Balmer emission lines of LRDs.

Our results show that the stellar masses of LRDs are within cosmological limits when MIRI data is included. However, extreme conditions including high baryon conversion efficiencies, $>0.5$ in some cases, and very dense stellar mass distributions that are $1-2$ dex higher than local elliptical galaxies are implied. Models of the formation and evolution of LRDs will have to account for efficient star formation at early times and the evolution of such dense distribution to systems with lower density at lower redshift.

\subsection{AGN Interpretation}

\subsubsection{Black Hole Masses}
\label{sec:mbh}

\begin{figure}[!t]
	\centering
    \includegraphics[width=0.5\textwidth]{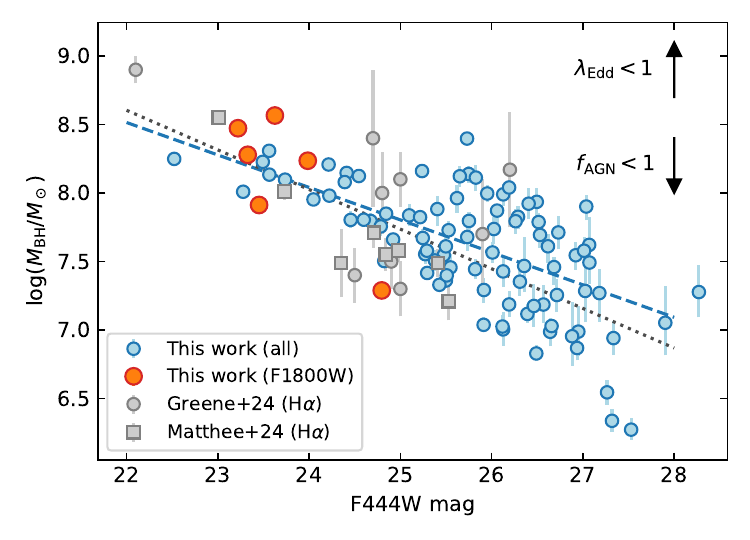}
		\caption{The mass of the central black hole assuming Eddington accretion versus F444W magnitude. We also show the results from spectroscopic observations of LRDs in \citet{greene24} and broad-line AGN in \citet{matthee24}. The best-fit linear relation between \mbh\ and F444W magnitude to the data in this study is shown in the blue dashed line, spectroscopic results in the grey dotted line. The upwawrd arrow indicates the direction in which $M_\mathrm{BH}$ would change if the Eddington ratio is decreased, while the downward arrow indicates that if the AGN contribution to the rest-optical is decreased. Our results are in agreement with those of spectroscopic studies. Given that our AGN bolometric luminosity are likely overestimated in the AGN-only model, the actual Eddington ratios are likely several times lower than unity.}
		\label{fig:mbh}
\end{figure}

We estimate the black hole mass (\mbh) from the bolometric AGN luminosity model assuming that the central black hole is accreting at Eddington luminosity, i.e. an Eddington ratio of unity. A lower Eddington ratio will lead to a higher \mbh\ than the estimate. Alternatively, a lower contribution of AGN emission will lead to a lower \mbh . Therefore, this estimate can be thought of as a lower limit to \mbh\ if the rest-optical continuum is solely produced by AGN emission. We show the derived \mbh\ versus the F444W magnitude in Figure \ref{fig:mbh}. For comparison, we also show \mbh\ measured in LRDs and broad-line AGNs in \citet{greene24} and \citet{matthee24} derived from virial relations using Balmer emission lines. We also plot the best-fit linear relation between \mbh\ and the F444W magnitude for our LRD sample and the spectroscopic results in the literature. 

The LRDs in our sample span \mbh\ of $\sim 10^{7-8} M_\odot$, which is comparable to those derived by virial relations. Both our results and the virial measurements increase with brighter F444W magnitude at a very similar rate, showing that our \mbh\ values are a reasonable estimate given their brightness in F444W. This apparent agreement can be a result of coincidence. Our \mbh\ estimates are based on two assumptions: an Eddington ratio of unity and a full AGN contribution to the red continuum. These two assumptions affect \mbh\ estimates in opposite directions, where the assumption of a lower Eddington ratio would increase \mbh , and a lower AGN contribution to the rest-optical would lead to a lower \mbh . We have shown in Section \ref{sec:res:agn} that the AGN bolometric luminosity is $\sim 0.6$ dex higher than spectroscopic results, suggesting an AGN contribution of less than unity to the red continuum. On the other hand, spectroscopic observations suggest that LRDs are accreting below the Eddington limit, with Eddington ratios spanning $\sim 0.1 - 0.4$ \citep{greene24}. Therefore, the apparent agreement between our \mbh\ and spectroscopic measurements, when combined with the higher inferred AGN bolometric luminosity in our results, could indicate that the central black holes in LRDs contribute only partly to the rest-optical continuum, and are accreting below the Eddington limit.

\subsubsection{Do LRDs host overmassive black holes?}

\begin{figure*}[thbp]
	\centering
    \includegraphics[width=0.9\textwidth]{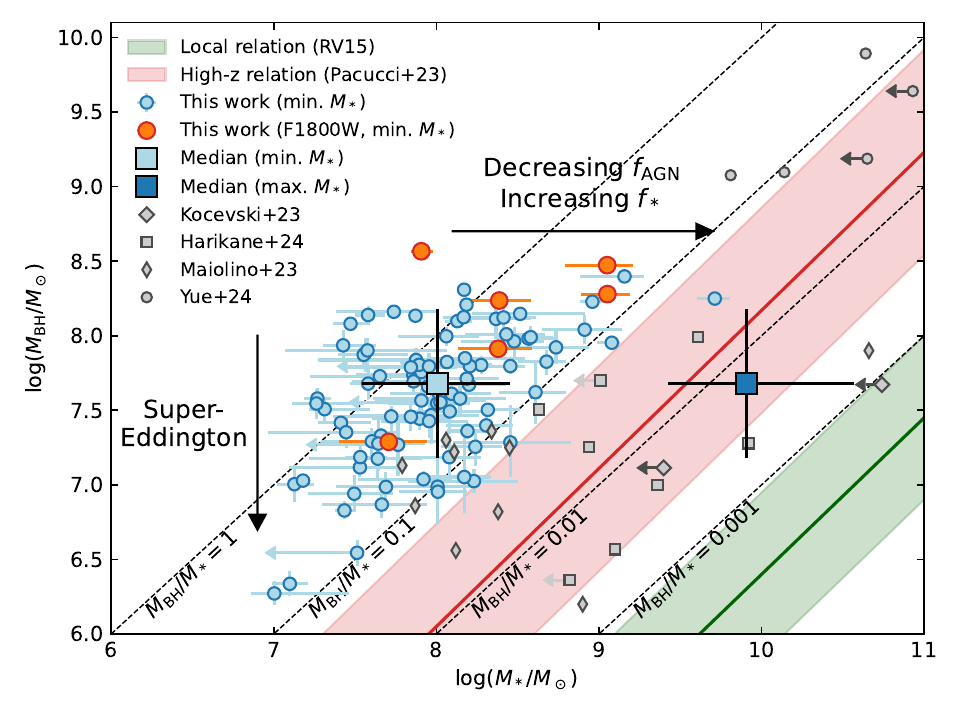}
		\caption{Black hole mass versus stellar mass. The LRDs in this study assuming the hybrid (minimal $M_*$) model are shown in the blue circles, while those detected in F1800W are shown in the orange circles. We also show the sample median for the minimal $M_*$ values in the light blue squares, as well as that for the maximal $M_*$ values from the galaxy-only fit in the dark blue squares. Literature values for high-redshift AGN observed by JWST from \citet{kocevski23, mai23, har23, yue24} are shown in the grey points. For context, we also plot the best-fit $\mbh - M_*$ relation and its scatter at high redshift in \citet{pac23} in the red line and shaded region, and the local universe in \citet{rv15} in blue. The $\mbh / M_*$ ratios inferred from the hybrid model are higher than the high-redshift relation by $\sim 1$ dex, while the galaxy-only model produces ratios at the lower end of the high-redshift relation. If the AGN contribution to the rest-optical continuum is less than unity, the stellar mass of the hybrid model can be underestimated, leading to an apparent elevation in $\mbh / M_*$ in the LRDs. Therefore, the rest-optical continuum of the LRDs are likely contributed by both AGN and galaxy emission. Alternatively, super-Eddington accretion could lower \mbh\ by $\sim 1$ dex, alleviating the tension in the $\mbh / M_*$ relation.}
		\label{fig:mbh_mstar}
\end{figure*}

Early JWST observations have suggested that SMBHs at $z > 4$ could violate the local relation between \mbh\ and the host galaxy stellar mass \citep{pac23, mai23, har23, kocevski23}, although selection biases can potentially produce a similar discrepancy \citep{li24}. For LRDs, studies using X-ray observations have argued that these objects do not host overmassive black holes, although the conclusion is dependent on the assumption of stellar mass estimates \citep{ananna24}. Here, we explore the relation between \mbh\ and the stellar mass of LRDs. In Figure \ref{fig:mbh_mstar}, we plot \mbh\ against the inferred stellar mass obtained from our hybrid SED model. We also show the best-fit relations at high redshift from \citet{pac23} and the local universe from \citet{rv15}.

Taking the \mbh\ and $M_*$ at face value, all the LRDs appear to host overmassive black holes. The typical $\mbh/M_*$ ratio of the LRDs ranges between $\sim 0.1-1$, about one dex more massive than AGN observed by JWST at similar redshifts \citep{kocevski23, mai23, har23, yue24}. Such masses are above the local relation by more than two dex, and are higher than even the high-redshift relation in \citet{pac23} by approximately one dex. For a fraction of the LRDs, the central black hole appears more massive than the host galaxy. This indicates that \mbh\ is overestimated and/or $M_*$ is underestimated in the LRDs. We argue that the former is less likely the explanation, since our \mbh\ estimates are in agreement with those measured by spectroscopy at a given F444W magnitude, if virial relations for \mbh\ are accurate at high redshift. Holding \mbh\ constant, $M_*$ will have to decrease to reconcile with existing observations. This suggests that $M_*$ in the hybrid model is likely underestimated. 

The hybrid model assumes that the rest-optical in the LRD SED is entirely produced by the AGN, while the stellar emission from the galaxy only contributes to the rest-UV. The inferred $M_*$ could be underestimated if the AGN does not produce the entirety of the rest-optical continuum, and some stellar light is responsible. In this sense, the hybrid model provides a ``minimum'' $M_*$ estimate in the LRDs. In fact, we have shown in the previous sections that the SED-derived AGN bolometric luminosity are systematically higher than spectroscopically-derived results in the literature, suggesting that the AGN contribution in the rest-optical is likely less than one. In this case, the stellar mass would be underestimated, leading to the elevated inferred $\mbh /M_*$ observed. 

To illustrate the effect, we also show in Figure \ref{fig:mbh_mstar} the sample medians for maximum and minimum stellar contribution in the rest-optical. For the minimum $M_*$ model, we adopt the \mbh\ and $M_*$ from the hybrid model, where the AGN contributes fully to the rest-optical, and the stellar emission only contributes to the rest-UV. For the maximum $M_*$ model, we replace the median $M_*$ with that from the galaxy-only model, such that stellar emission produces the entirety of the rest-optical. We keep the median \mbh\ unchanged from the hybrid model, since our \mbh\ are in good agreement with those determined by spectroscopy using virial relations (see Section \ref{sec:mbh}). \mbh\ can remain constant with increasing stellar and decreasing AGN contributions if the Eddington ratio decreases from unity. 

Another possibility to reconcile the \mbh -$M_*$\ relation is that the central black holes in LRDs are undergoing super-Eddington accretion. Our \mbh\ estimates assume an Eddington ratio of unity, and provides a lower limit under normal sub-Eddington scenarios. Under super-Eddington accretion, a lower \mbh\ is required to produce the same observed luminosity, reducing the observed \mbh -$M_*$ ratios. \citet{lambrides24} have shown that super-Eddington accretion could potentially explain the X-ray weakness of LRDs with an optically thick accretion disk preventing UV photons from being scattered to high energy. They have also found that the inferred \mbh\ under super-Eddington accretion would be reduced by at least an order of magnitude compared with those measured from Balmer emission lines. The tension with the \mbh -$M_*$\ relation in our results can be alleviated if the \mbh\ are reduced by a similar amount of at least one dex.

\subsubsection{The lack of hot dust signature and the ubiquity of broad Balmer emission lines}

An intriguing feature of the LRD SED is their weak mid-IR emission \citep{williams23, pg23}, which results in very low \fhd\ in our AGN models. Here, we discuss several possible physical scenarios leading to the observed mid-IR weakness if LRDs are indeed powered by AGNs.

First, based on our derived AGN bolometric luminositiy and \mbh\ to $M_*$ ratios, it is likely that the AGN contribution to the rest-optical is less than unity. In fact, our SED-derived AGN bolometric luminosities are $\sim 0.6$ dex lower than, i.e. $\sim 1/4$ times of, those derived from spectroscopy. Consequently, the corresponding hot dust emission in the mid-IR will be lower than expected when common AGN templates, which assume full AGN contribution,  are used to predict the mid-IR emission in LRDs. Similarly, \fhd\ in our SED modeling assumes that the entire rest-optical emission is produced by the AGN, potentially underestimating \fhd\ by a factor of $\sim 5$. The median \fhd\ for our LRDs brighter than 26 mag in F444W is $\sim 0.04$. Even after scaling by this correction factor, the median \fhd\ of $\sim 0.2$ remains substantially below unity. While this overestimated AGN bolometric luminosity partially mitigates the discrepancy, it is insufficient to fully explain the mid-IR weakness.

In the conventional unified AGN model, a region of the dusty molecular gas---known as the dusty torus---exists in the close proximity ($\sim 0.1-10$ pc) of the accretion disk. This dusty torus, heated to the sublimation temperature of the dust grains ($\sim 1000-1500$ K), obscures UV/visible photons from the accretion disk and BLR from some sight lines, reprocessing this energy into thermal IR emission. In most obscured AGN, the obscuration is thought to occur in this dusty structure in close proximity of the central black hole \citep[e.g.][]{hickox18}. In our SED modeling results, we find that the amount of obscuring, as measured by $A_V$, is uncorrelated with the amount of hot dust, as measured by \fhd ,  challenging the idea that a conventional torus at $\sim 1000$ K is responsible for the obscuration in these AGNs.

In fact, a broader variety of dust distribution around the AGN can also produce obscuration. Extended polar dust, associated with the narrow-line region ($\sim 10-1000$ pc from the central black hole), has been observed in nearby AGNs \citep{lyu18}. In addition, obscuration can also come from the host galaxy, as observed in edge-on galaxies in the local universe \citep[e.g.][]{goulding12, buchner17}. These more extended dust distributions are subject to less intense heating from the accretion disk, leading to lower dust temperatures. Recently, \citet{li24b} showed that in the absence of a centrally concentrated torus, the IR emission at $\sim 3 \mu$m can be suppressed if the dust around the central engine resides in a more extended dust density profile. This distribution can be produced during an early phase of the evolution of the galactic nuclei, as dust grains produced in the host galaxy accrete onto the nucleus before a well-defined dust torus structure forms. Interestingly, the ubiquity of broad Balmer line detections \citep[e.g.][]{greene24, kocevski24, taylor24} is consistent with a more isotropic but less optically thick obscuring dusty structure than a well-defined torus. Before JWST, at least some AGNs are known to be deficient in hot dust. This hot-dust-deficient population exists from the local universe up to $z=6$ \citep{hao10, jiang10, lyu17}, and it has been reported that the fraction of hot-dust-deficient AGN increases with redshift from $z\sim 0$ to 4. Given the sharp increase in abundance for LRDs at $z \gtrsim 4$ \citep{kocevski24}, they could represent an early phase of SMBH formation where the torus is being formed.

\subsection{The Origin of the Red Continuum}

In the previous subsections, we have examined the implications of models which assume the red continuum is dominated by either purely AGN or stellar emission. Both will lead to extreme physical properties in LRDs which are challenging to interpret. While the galaxy-only model is consistent with the observed Balmer breaks in some LRDs, and produces stellar masses that are not ruled out by current cosmological models, it leads to extreme stellar mass surface densities that are $1-4\sigma$ higher than those in typical galaxies at similar redshifts and $1-2$ dex higher than local elliptical galaxies, requiring a drastic evolution in their stellar mass distribution to produce descendants similar to present-day galaxies. For the hybrid model, which assumes AGN domination in the red continuum, energetic considerations implies black hole masses that are more extreme than even the elevated $M_\mathrm{BH}-M_*$ relation recently reported at high redshift. With these consideration, it could be reasonable to conclude that there is a mixture of stellar and AGN contribution to the rest-frame optical continuum. This will lead to intermediate stellar and/or black hole masses that are between the two extreme models examined, partially alleviating both the issue of the high stellar mass surface density and overmassive black holes. Alternatively, the presence of super-Eddington accretion can also lead to lower black hole and stellar masses than estimated by the conventional galaxy and AGN models in this study.

If the AGN luminosity is indeed overestimated by $\sim 0.6$ dex as shown in the previous sections, this would imply that the AGN contributes to $\sim 25\%$ to the rest-optical continuum, while stellar emission accounts for $\sim 75\%$. While such stellar masses could reconcile the extreme $M_\mathrm{BH}-M_*$ relation, they are likely still too high to be consistent with the observed galaxy size-mass relation. Determining the exact contribution of galaxy and AGN in LRDs is challenging with only broadband photometry available in this study, where important features such as the Balmer break and continuum slopes are not resolved. Some recent work have used deep spectroscopy with the help of lensing magnification, capable of detecting these features, to model LRD continuum emission and found AGN contribution of $\sim 0.1-0.2$ in the rest-optical continuum in a triply-lensed LRD \citep{ma24}, although similar challenges in interpreting the model are also noted by the authors. In the future, expanding the sample of LRDs with deep spectroscopy to fully characterize the continuum can assess the prevalence of features such as Balmer breaks, and determine stellar contribution across the population.

\section{Conclusions} \label{sec:conclusions}

We use data from the PRIMER survey to analyze the SED of 95 LRDs, the largest sample to-date with full photometric coverage throughout $\sim 1-18~\mu$m. We examine three models representing the idealized scenario where either galaxy or AGN emission dominates the rest-frame UV or optical continuum. We measure physical properties from each model and explore their respective consequences. Our main findings are summarized as follows.

\begin{enumerate}
    \item The galaxy-only model results in a dusty, massive ($\sim 10^{10}~M_\odot$) stellar populations in the rest-frame optical and an unobscured, low-mass ($\sim 10^{8}~M_\odot$) component in the rest-frame UV. Both components have similar ages, which could be produced by inhomogeneous obscuration of a single stellar population.

    \item The AGN-only model produces dusty luminous AGNs, with bolometric luminosities of $10^{45-46.5}~\mathrm{erg~s}^{-1}$, and dust attenuations of $A_V \sim 2-4$. The hot dust content of these AGNs are typically $\lesssim 0.2$ of that in common quasars.

    \item The hybrid model produces a low-mass  ($\sim 10^{8}\ M_\odot$) unobscured galaxy in the rest-frame UV. The stellar mass of LRDs are thus uncertain at $\sim 2$ dex depending on the model selected.

    \item With the stellar interpretation, the inclusion of MIRI photometry reduces stellar masses of $z\gtrsim 7$ LRDs by $\sim 0.4$ dex, producing values that do not violate cosmological models, but a very high efficiency baryon conversion efficiency ($\epsilon = 0.2-1$) is still required in a subset of LRDs. Extreme stellar mass densities are at least $> 1-4\sigma$ higher than the size-mass relation at similar redshifts and $1-2$ dex higher than local elliptical galaxies.

    \item With the AGN interpretation, the AGN bolometric luminosities measured in our SED fitting are higher than that from spectroscopic measurements in LRDs at the same F444W brightness, but the black hole mass assuming Eddington accretion are similar to spectroscopic results. This suggests that the AGN may not dominate the rest-frame optical, and the central black hole is accreting at sub-Eddington rates. 
    
    \item The black hole and stellar mass estimates from the hybrid model result in highly overmassive black holes above other high-redshift AGN. This suggests that some stellar emission is responsible for the rest-frame optical, or super-Eddington accretion is leading to overestimated black hole mass estimates.

\end{enumerate}

Our modeling shows that extreme physical conditions are required in either scenarios where the AGN or galaxy dominates the red continuum in LRDs. This can be taken as evidence of a mixture of contribution from both sources, which can alleviate at least some of the tension observed. The direct characterization of the continuum is not possible given the low spectral resolution of broadband photometry, and deep spectroscopy of the LRD continuum and variability monitoring in the future can provide more insight into the origin of the red continuum in the characteristic LRD SED.

\begin{acknowledgments}

We thank Jenny Greene for useful discussions.
We acknowledge the RUBIES collaboration for designing and executing their observations with a zero-exclusive-access period, from which we derived spectroscopic redshifts for nine of the LRDs in our sample.
This work is based on observations made with the NASA/ESA/CSA James Webb Space Telescope. The data were obtained from the Mikulski Archive for Space Telescopes at the Space Telescope Science Institute, which is operated by the Association of Universities for Research in Astronomy, Inc., under NASA contract NAS 5-03127 for JWST. These observations are associated with programs \#1837 and \#4233.
G.C.K.L. and S.L.F. acknowledge support from NASA/STScI through award JWST-GO-1837.
R.S.E acknowledges generous financial support from the Peter and Patricia Gruber Foundation.
D.J.M acknowledges the support of the Science and Technology Facilities Council. J.S.D. acknowledges the support of the Royal Society via the award of a Royal Society Research Professorship.

G.C.K.L., S.L.F., A.M.M., A.J.T., H.B.A., O.A.C.O. and V.K. acknowledge that the location where they work, the University of Texas at Austin, sits on indigenous land. The Tonkawa lived in central Texas, and the Comanche and Apache moved through this area. We pay our respects to all the American Indian and Indigenous Peoples and communities who have been or have become a part of these lands and territories in Texas, on this piece of Turtle Island.

\end{acknowledgments}

\vspace{5mm}
\facilities{JWST(NIRCam and MIRI), HST(ACS)}

\software{\texttt{astropy} \citep{2013A&A...558A..33A,2018AJ....156..123A},  
          \textsc{Bagpipes} \citep{car18},
          \texttt{EAZY} \citep{brammer08}
          \textsc{galfit} \citep{peng2002},
          \texttt{Source Extractor} \citep{1996A&AS..117..393B}, 
          }

\appendix

\section{Tabulated Data}

We provide tabulated data of the sample and its derived physical properties. In Table \ref{tab:results}, we show a sample of the table, the entirety of which is available in machine-readable form in the online journal.

\begin{longrotatetable} 
\begin{deluxetable*}{lrrrrrrrrrrrrrr}
%\vspace{2mm}
%\tabletypesize{\small}
\tablecaption{SED results}
\tablewidth{\textwidth}
\tabletypesize{\footnotesize}
\tablehead{\multicolumn{7}{c}{} & \multicolumn{3}{c}{Galaxy-only Model} & \multicolumn{3}{c}{AGN-only Model} & \multicolumn{2}{c}{Hybrid Model}
\\
\cmidrule(lr){8-10} \cmidrule(lr){11-13} \cmidrule(lr){14-15}
\multicolumn{1}{l}{ID} & \multicolumn{1}{c}{R.A.} & \multicolumn{1}{c}{Decl.} & \multicolumn{1}{c}{$z_\mathrm{phot}$} & \multicolumn{1}{c}{$z_\mathrm{spec}$} & \multicolumn{1}{c}{F770W\tablenotemark{a}} & \multicolumn{1}{c}{F1800W\tablenotemark{a}} & \multicolumn{1}{c}{$M_*$} & \multicolumn{1}{c}{$A_{V,1}$} & \multicolumn{1}{c}{$A_{V,2}$} & \multicolumn{1}{c}{$\log(L_\mathrm{bol})$} & \multicolumn{1}{c}{$A_V$} & \multicolumn{1}{c}{\fhd} & \multicolumn{1}{c}{$M_*$} & \multicolumn{1}{c}{$A_V$}
\\
\multicolumn{1}{c}{} & \multicolumn{1}{c}{[deg]} & \multicolumn{1}{c}{[deg]} & \multicolumn{2}{c}{} & \multicolumn{1}{c}{[AB mag]} & \multicolumn{1}{c}{[AB mag]} & \multicolumn{1}{c}{[$M_\odot~\mathrm{yr}^{-1}$]} & \multicolumn{1}{c}{[mag]} & \multicolumn{1}{c}{[mag]} & \multicolumn{1}{c}{[$\mathrm{erg~s}^{-1}$]} & \multicolumn{1}{c}{[mag]} & \multicolumn{1}{c}{$(\times 10^{-3})$} & \multicolumn{1}{c}{[$M_\odot~\mathrm{yr}^{-1}$]} & \multicolumn{1}{c}{[mag]}
}
\startdata
COS-2144 & 150.13991 & 2.44250 & 4.66 & - & $<25.55$ & $<23.40$ & $8.6^{+0.1}_{-0.1}$ & - & $1.0^{+0.1}_{-0.1}$ & $44.4^{+0.1}_{-0.1}$ & $1.8^{+0.3}_{-0.3}$ & $684.8^{+549.3}_{-490.4}$ & $7.0^{+0.5}_{-0.1}$ & $0.3^{+0.2}_{-0.1}$\\
COS-2356 & 150.16342 & 2.43916 & 5.14 & - & 24.16 & $<23.36$ & $10.5^{+0.2}_{-0.2}$ & $1.1^{+0.5}_{-0.7}$ & $3.4^{+0.4}_{-0.3}$ & $46.1^{+0.0}_{-0.0}$ & $3.0^{+0.1}_{-0.1}$ & $12.7^{+12.0}_{-9.3}$ & $8.6^{+0.2}_{-0.2}$ & $0.3^{+0.2}_{-0.2}$\\
COS-2451 & 150.17328 & 2.43768 & 4.69 & - & 24.13 & $<22.97$ & $9.9^{+0.1}_{-0.1}$ & - & $1.3^{+0.1}_{-0.1}$ & $45.6^{+0.0}_{-0.0}$ & $2.1^{+0.2}_{-0.1}$ & $152.1^{+30.2}_{-43.0}$ & $8.3^{+0.2}_{-0.0}$ & $1.0^{+0.0}_{-0.0}$\\
COS-5229 & 150.15525 & 2.40667 & 4.72 & - & 25.06 & $<23.83$ & $9.5^{+0.1}_{-0.1}$ & - & $2.5^{+0.2}_{-0.2}$ & $45.2^{+0.1}_{-0.1}$ & $2.7^{+0.2}_{-0.2}$ & $215.1^{+46.9}_{-64.3}$ & $7.5^{+0.3}_{-0.4}$ & $0.3^{+0.5}_{-0.2}$\\
COS-5621 & 150.15583 & 2.40311 & 5.23 & - & 25.43 & $<22.80$ & $9.7^{+0.1}_{-0.1}$ & - & $1.9^{+0.1}_{-0.1}$ & $45.5^{+0.0}_{-0.0}$ & $2.7^{+0.1}_{-0.1}$ & $43.4^{+71.4}_{-32.6}$ & $7.4^{+0.2}_{-0.1}$ & $0.4^{+0.1}_{-0.1}$\\
COS-5844 & 150.13710 & 2.40121 & 8.2 & - & 25.56 & $<23.25$ & $9.9^{+0.2}_{-0.1}$ & $0.1^{+0.2}_{-0.1}$ & $3.8^{+0.2}_{-0.1}$ & $46.1^{+0.1}_{-0.1}$ & $3.3^{+0.2}_{-0.2}$ & $154.0^{+110.2}_{-111.0}$ & $8.6^{+0.1}_{-0.2}$ & $0.1^{+0.1}_{-0.1}$\\
COS-7071 & 150.14236 & 2.39002 & 5.23 & - & 24.33 & $<23.07$ & $10.6^{+0.1}_{-0.1}$ & $1.3^{+0.2}_{-0.3}$ & $6.3^{+0.3}_{-0.4}$ & $45.8^{+0.1}_{-0.1}$ & $4.2^{+0.3}_{-0.3}$ & $126.1^{+63.1}_{-59.0}$ & $7.9^{+0.2}_{-0.4}$ & $0.2^{+0.2}_{-0.1}$\\
COS-7236 & 150.17017 & 2.38839 & 5.29 & - & 23.77 & $<23.12$ & $9.9^{+0.1}_{-0.1}$ & $0.1^{+0.1}_{-0.1}$ & $1.5^{+0.0}_{-0.0}$ & $46.2^{+0.0}_{-0.0}$ & $3.0^{+0.1}_{-0.1}$ & $11.7^{+10.2}_{-8.4}$ & $8.1^{+0.1}_{-0.1}$ & $0.3^{+0.1}_{-0.1}$\\
COS-8227 & 150.15684 & 2.37946 & 6.67 & - & 24.90 & $<23.43$ & $10.7^{+0.1}_{-0.1}$ & $1.8^{+0.1}_{-0.4}$ & $5.1^{+0.3}_{-0.4}$ & $46.1^{+0.1}_{-0.1}$ & $3.6^{+0.1}_{-0.1}$ & $43.4^{+33.6}_{-29.5}$ & $8.1^{+0.2}_{-0.3}$ & $0.2^{+0.2}_{-0.2}$\\
COS-8406 & 150.15046 & 2.37787 & 5.17 & - & 25.37 & $<23.48$ & $9.7^{+0.1}_{-0.1}$ & - & $3.1^{+0.2}_{-0.2}$ & $45.5^{+0.1}_{-0.1}$ & $3.5^{+0.3}_{-0.3}$ & $203.1^{+93.8}_{-96.5}$ & $7.4^{+0.3}_{-0.5}$ & $0.2^{+0.4}_{-0.2}$\\
COS-8550 & 150.17590 & 2.37676 & 4.93 & - & 25.44 & $<23.36$ & $9.8^{+0.1}_{-0.1}$ & - & $2.9^{+0.2}_{-0.4}$ & $45.5^{+0.1}_{-0.1}$ & $3.4^{+0.2}_{-0.2}$ & $90.3^{+93.0}_{-66.3}$ & $7.9^{+0.2}_{-0.3}$ & $0.1^{+0.2}_{-0.1}$\\
COS-11625 & 150.06333 & 2.35498 & 4.72 & - & 24.17 & $<23.32$ & $9.7^{+0.1}_{-0.1}$ & - & $1.6^{+0.1}_{-0.2}$ & $45.5^{+0.1}_{-0.1}$ & $2.5^{+0.2}_{-0.2}$ & $184.3^{+34.5}_{-37.1}$ & $8.2^{+0.1}_{-0.1}$ & $1.2^{+0.1}_{-0.1}$\\
COS-12184 & 150.07671 & 2.35092 & 5.17 & - & 25.46 & $<22.87$ & $9.1^{+0.1}_{-0.1}$ & - & $1.9^{+0.2}_{-0.2}$ & $45.1^{+0.1}_{-0.1}$ & $2.8^{+0.3}_{-0.3}$ & $944.9^{+302.2}_{-292.0}$ & $7.1^{+0.1}_{-0.1}$ & $0.3^{+0.2}_{-0.1}$\\
COS-13104 & 150.14444 & 2.34440 & 4.54 & - & 23.97 & $<23.35$ & $10.4^{+0.0}_{-0.1}$ & $0.0^{+0.0}_{-0.0}$ & $2.0^{+0.2}_{-0.1}$ & $46.1^{+0.0}_{-0.0}$ & $2.9^{+0.1}_{-0.1}$ & $5.6^{+7.0}_{-4.2}$ & $9.1^{+0.1}_{-0.1}$ & $0.1^{+0.1}_{-0.0}$\\
COS-13886 & 150.17577 & 2.33879 & 5.86 & - & 23.27 & $<23.11$ & $10.7^{+0.1}_{-0.1}$ & $0.1^{+0.1}_{-0.1}$ & $2.5^{+0.1}_{-0.1}$ & $46.4^{+0.0}_{-0.0}$ & $2.9^{+0.1}_{-0.1}$ & $1.6^{+2.4}_{-1.2}$ & $8.2^{+0.0}_{-0.0}$ & $0.4^{+0.0}_{-0.0}$\\
COS-15702 & 150.07568 & 2.32529 & 4.69 & - & $<26.09$ & $<23.25$ & $8.5^{+0.1}_{-0.1}$ & - & $1.1^{+0.1}_{-0.1}$ & $44.4^{+0.1}_{-0.1}$ & $1.8^{+0.3}_{-0.3}$ & $756.1^{+520.7}_{-500.5}$ & $7.1^{+0.1}_{-0.1}$ & $0.4^{+0.2}_{-0.1}$\\
COS-18805 & 150.06783 & 2.30330 & 6.43 & - & 24.68 & $<23.00$ & $10.2^{+0.1}_{-0.1}$ & $0.4^{+0.6}_{-0.3}$ & $3.1^{+0.3}_{-0.4}$ & $46.1^{+0.1}_{-0.1}$ & $3.6^{+0.2}_{-0.3}$ & $78.7^{+68.1}_{-55.4}$ & $8.5^{+0.2}_{-0.3}$ & $0.2^{+0.2}_{-0.2}$\\
COS-20188 & 150.07603 & 2.29377 & 5.77 & - & 25.01 & $<23.49$ & $10.0^{+0.1}_{-0.1}$ & - & $2.2^{+0.1}_{-0.1}$ & $45.8^{+0.1}_{-0.1}$ & $2.8^{+0.2}_{-0.2}$ & $53.4^{+52.9}_{-38.2}$ & $8.2^{+0.2}_{-0.3}$ & $0.2^{+0.3}_{-0.1}$\\
COS-20492 & 150.05536 & 2.29159 & 6.43 & - & 24.54 & $<23.50$ & $9.8^{+0.1}_{-0.1}$ & - & $1.1^{+0.1}_{-0.1}$ & $45.9^{+0.1}_{-0.1}$ & $2.5^{+0.2}_{-0.1}$ & $46.7^{+35.2}_{-33.3}$ & $8.3^{+0.1}_{-0.1}$ & $0.3^{+0.1}_{-0.1}$\\
COS-20652 & 150.07852 & 2.29044 & 4.69 & - & 25.25 & $<23.47$ & $9.5^{+0.1}_{-0.1}$ & - & $2.1^{+0.3}_{-0.3}$ & $45.1^{+0.1}_{-0.1}$ & $2.3^{+0.3}_{-0.3}$ & $351.7^{+95.6}_{-106.4}$ & $8.2^{+0.3}_{-0.4}$ & $1.0^{+0.3}_{-0.5}$\\
COS-21273 & 150.07916 & 2.28573 & 6.37 & - & 24.96 & $<23.48$ & $9.9^{+0.1}_{-0.1}$ & $0.1^{+0.1}_{-0.1}$ & $3.0^{+0.2}_{-0.2}$ & $45.8^{+0.1}_{-0.1}$ & $3.3^{+0.2}_{-0.2}$ & $98.5^{+79.2}_{-67.2}$ & $7.6^{+0.4}_{-0.1}$ & $0.2^{+0.1}_{-0.1}$\\
COS-21309 & 150.11153 & 2.28552 & 5.23 & - & 26.15 & $<23.19$ & $9.3^{+0.1}_{-0.1}$ & - & $1.9^{+0.2}_{-0.2}$ & $45.1^{+0.1}_{-0.1}$ & $2.7^{+0.3}_{-0.3}$ & $366.1^{+354.9}_{-242.4}$ & $7.7^{+0.2}_{-0.3}$ & $0.2^{+0.2}_{-0.1}$\\
COS-22424 & 150.08281 & 2.27772 & 5.35 & - & 23.86 & $<23.30$ & $10.5^{+0.1}_{-0.1}$ & $0.2^{+0.2}_{-0.1}$ & $4.3^{+0.2}_{-0.2}$ & $46.3^{+0.0}_{-0.0}$ & $4.6^{+0.0}_{-0.1}$ & $13.7^{+9.8}_{-8.5}$ & $7.7^{+0.3}_{-0.4}$ & $0.2^{+0.2}_{-0.1}$\\
COS-22433 & 150.07802 & 2.27762 & 6.28 & - & 24.04 & $<23.56$ & $10.6^{+0.0}_{-0.1}$ & $0.1^{+0.2}_{-0.1}$ & $2.8^{+0.1}_{-0.1}$ & $46.2^{+0.0}_{-0.0}$ & $3.1^{+0.1}_{-0.1}$ & $3.8^{+5.5}_{-2.7}$ & $8.5^{+0.4}_{-0.3}$ & $0.8^{+0.1}_{-0.1}$\\
COS-25273 & 150.08470 & 2.25401 & 8.56 & - & 25.82 & $<23.45$ & $10.1^{+0.1}_{-0.1}$ & $0.3^{+0.2}_{-0.2}$ & $4.0^{+0.2}_{-0.2}$ & $45.9^{+0.1}_{-0.1}$ & $3.0^{+0.2}_{-0.2}$ & $214.8^{+163.5}_{-149.2}$ & $8.7^{+0.1}_{-0.2}$ & $0.2^{+0.1}_{-0.1}$\\
\enddata
\tablenotetext{a}{5$\sigma$ upper limits are given for non-detections, and are denoted as negative numbers in the machine-readable table.}
\tablecomments{This table is available in its entirety in machine-readable form.}
\label{tab:results}
%\vspace{-8mm}
\end{deluxetable*}
\end{longrotatetable}

\bibliography{mybib}{}
\bibliographystyle{aasjournal}

\end{document}